# High magnetic anisotropy and magnetocaloric effects in single crystal Cr$_2$Te$_3$

Anirban Goswami[1], Nicholas Ng[2,3], AM Milinda Abeykoon[4], Emmanuel Yakubu[1] and Samaresh Guchhait[1*]

1. Department of Physics and Astronomy, Howard University, Washington, DC 20059, USA
2. Department of Chemistry, The Johns Hopkins University, Baltimore, MD 21218, USA
3. Institute for Quantum Matter, The William H. Miller III Department of Physics and Astronomy, The Johns Hopkins University, Baltimore, MD 21218, USA
4. National Synchrotron Light Source-II, Brookhaven National Laboratory, Upton, NY 11733, USA

## Abstract

We report a systematic investigation of anisotropic magnetocaloric effects in single crystal Cr$_2$Te$_3$. Single crystal samples are synthesized by chemical vapor transport and characterized by x-ray and Laue diffraction methods. The maximum magnetic entropy change $-\Delta S_M^{max}$ is 4.50 J kg$^{-1}$ K$^{-1}$ for the easy *c*-axis (3.36 J kg$^{-1}$ K$^{-1}$ for the hard *ab*-plane) and the relative cooling power RCP is 296.7 J kg$^{-1}$ for the easy *c*-axis (183.84 J kg$^{-1}$ for the hard axis *ab*-plane) near the Curie temperature for a magnetic field change of 9 T. The magneto-crystalline anisotropy constant K$_u$ is estimated to be 486.92 kJ m$^{-3}$ at 146 K, decreasing to 148.60 kJ m$^{-3}$ at 168 K. Meanwhile, the maximum of the rotational magnetic entropy change $\Delta S_M^R$(T, H) between the *c*-axis and the *ab*-plane is about 1.14 J kg$^{-1}$ K$^{-1}$ for magnetic-field change of 9 T. The critical exponents are estimated by analyzing magnetocaloric effects, which indicate 2D-Ising type magnetic system. The accuracy of estimated critical exponents is verified by scaling analysis. The maximum magnetic entropy change $-\Delta S_M^{max}$ ~5.25 J kg$^{-1}$ K$^{-1}$ (along the *c*-axis) and the corresponding adiabatic temperature change $\Delta T_{ad}$ ~3.31 K (along the *c*-axis) are estimated by analyzing heat capacity measurements with a magnetic field up to 9 T.

## I. Introduction

After the discovery of long-range intrinsic magnetism down to the monolayer limit in two-dimensional materials, transition-metal chalcogenides have received renewed interest from the academic community because some of these binary phases may show room temperature ferromagnetism down to monolayer limit which could be suitable for low dimensional spintronics applications [1]. Ultrathin layers of transition metal chalcogenides show great potential in device applications owing to the tunable correlation between their structure and their physical and electrical properties [2–7]. Among these transition metal chalcogenides, chromium telluride ($Cr_xTe_y$) binary phases have gained significant attention because of their potential applications in spintronics. Previously, many binary phases of $Cr_xTe_y$ (such as, CrTe [8], $CrTe_2$ [9,10], $Cr_2Te_3$ [11–14], $Cr_3Te_4$ [15], $Cr_5Te_6$ [16], $Cr_{0.62}Te$ [17], $Cr_5Te_8$ [18,19]) have been studied to explore their structure-property relationships. Some of its compositions exhibit room-temperature ferromagnetic properties in bulk and layered forms [8,15,16,20]. These compounds are reported to be ferromagnets with Curie temperatures ($T_C$) ranging from 160 to 340 K, depending on their chromium concentration [21]. There are alternating stacks of Cr-deficient and Cr-full layers along the *c*-axis in these compounds, and the vacancy in the chromium content plays an important role in determining their structure and magnetic properties [22].

In this article, we will report the anisotropic magnetocaloric properties of single crystal $Cr_2Te_3$. The magnetic properties of bulk $Cr_2Te_3$ and the correlation between their structural and magnetic properties have been studied before [23–26]. Recently, there have been some studies on $Cr_2Te_3$ in both bulk and layered forms [11–14]. The neutron diffraction study shows ferromagnetic ordering in $Cr_2Te_3$ along the *c*-axis and a reduction in the volume of the unit cell with decreasing temperature [25,27]. The average calculated magnetic moment is 2.7 $\mu_B$/Cr atom in this material, which is significantly smaller than that of the $Cr^{3+}$ ion [28]. According to John Goodenough [29], this reduction in the effective magnetic moment may originate from a spiral antiferromagnetic spin structure, whereas an alternative mechanism suggested by A. F. Anderson *et al*. [28] indicates that this phenomenon may be due to magnetic moment canting. Yao Wen *et al.* reported that the Curie temperature of $Cr_2Te_3$ can be tuned near the room temperature (~280 K) by decreasing the thickness up to 1 or 2 unit-cell which is supported by the anomalous Hall effect measurement [14]. Fang Wang *et al.* reported the organic solution phase synthesis of $Cr_2Te_3$ nanorods with ultra-high coercivity [11]. Roy *et al.* have grown $Cr_2Te_3$ thin films by molecular beam epitaxy which show large perpendicular magnetic anisotropy [12,13].

However, our goal here is to explore the anisotropic magnetocaloric properties of single crystal $Cr_2Te_3$ which has not been reported before. Magnetic refrigeration using the magnetocaloric effect (MCE) has a huge potential to meet the worldwide demand for environmentally friendly, green, and energy-efficient thermal management [30,31]. As $Cr_2Te_3$ exhibits a second-order magnetic phase transition, it exhibits a change in magnetic entropy over a broad temperature range, and it does not exhibit magnetic or thermal hysteresis. Due to the 2$^{nd}$ order magnetic phase transition, it may be suitable for potential magnetocaloric material applications because there is a continuous change from the low-temperature phase to the high-temperature phase. In contrast, in the 1$^{st}$ order magnetic phase transition there is a temperature range where both phases coexist [32]. Till now, anisotropic magnetocaloric properties have been reported in other layered magnetic materials, such as $CrI_3$ [33], $CrCl_3$ [34,35], $CrSbSe_3$ [36], and $Cr_4Te_5$ [20]. There are reports on the estimation of critical exponents by analyzing the anisotropic magnetocaloric effects in some materials, such as $Cr_2Ge_2Te_6$ [37] and $Cr_5Te_8$ [19].

Here, we report the anisotropic magnetocaloric properties of a single crystal $Cr_2Te_3$ sample by studying field-dependent isothermal magnetization along its *c*-axis and *ab*-plane. By analyzing the magneto-caloric effects, we show there is a 2$^{nd}$ order (paramagnetic to ferromagnetic) magnetic phase transition. Moreover, by determining the magneto-caloric effects as a function of field and temperature, we calculate its full width at half maximum (FWHM), and refrigerant capacity or relative cooling power (RCP). With the help of these physical parameters, the critical exponents related to this magnetic phase transition are estimated. We then verify the accuracy of our estimated critical exponents by scaling analysis. Lastly, we estimated adiabatic temperature and change of magnetic entropy from the temperature-dependent heat capacity measurement to justify our analysis.

## II. Sample Synthesis and Phase Characterizations

Single crystal samples were synthesized by the chemical vapor transport (CVT) method. These CVT experiments were performed in a Thermo Scientific Lindberg Blue M three-zone furnace equipped with UP150 model program controllers. Cr (Alfa Aesar, powder, -100+325 mesh, 99.99% metals basis) and Te (Thermo Scientific, shot, 2-5 mm diameter, 99.9999% metals basis) were used as received without further purification. Stoichiometric amounts of Cr and Te are combined in a fused quartz tube. Between 35-50 mg of solid iodine is used as the vapor transport agent. The iodine is added to the tube, which is then sealed under vacuum. The sealed tube is, at minimum, long enough to equal the distance between two zones of a three-zone furnace. This tube is then placed inside the three-zone furnace. All three zones are heated up at a rate of 100°C per hour, with the charge zone reaching 900°C and the crystallization zone reaching 800°C. This temperature gradient is held for one week, then the furnace is cooled at a rate of 100°C per hour to room temperature. Room temperature x-ray powder diffraction measurements were carried out at the Pair Distribution Function (PDF) beamline (28-ID-1) of the National Synchrotron Light Source-II. We collected our data on grounded single crystal samples in capillary transmission geometry using a PerkinElmer amorphous silicon detector placed 1000 mm downstream from the sample [17]. The setup utilized a 74.5 keV ($\lambda$ = 0.1665 Å) x-ray beam. Two-dimensional diffraction images were radially integrated to obtain intensity vs. 2-theta data using the pyFAI software package. The Rietveld refinement was carried out using the GSAS-II software package [38]. Figure 1(a) shows the Rietveld fit to experimental data using the symmetry space group $P\bar{3}1c$. This is a different space group from the reported $Cr_{0.62}Te$ phase symmetry space group $P\bar{3}m1$ which is also the same as the symmetry space group of the reported $Cr_5Te_8$ phase [17,39]. The refined unit cell parameters are a, b = 6.7920(2) Å, and c = 12.0887(2) Å. The refined fractional coordinates, occupancy of Cr and Te sites, and the isotropic thermal displacement parameters ($U_{iso}$) are shown in Table I. Figure 1(b) shows the refined crystal structure of trigonal $Cr_{1.77}Te_3$ generated using the Vesta software package [40]. The 2c (0.333, 0.666, 0.25) sites that show Cr deficiency are shown with half-white half-blue balls in Fig. 1(b). The Laue diffraction is used to determine the crystallographic axes of a single crystal sample, as shown in Figure 1(c).

Table I: Results of the Rietveld refinement.

| Atom | Site | x | y | z | Occupancy | Uiso (Å$^2$) |
|---|---|---|---|---|---|---|
| Cr | 2c | 0.3333 | 0.6667 | 0.537(1) | 0.5370(1) | 0.011(5) |
| Cr | 2b | 0.0000 | 0.0000 | 0.0000 | 1.0000 | 0.007(8) |
| Cr | 4f | 0.3333 | 0.6667 | -0.0032(8) | 1.0000 | 0.005(4) |
| Te | 12i | 0.3351(7) | 0.0019(5) | 0.3732(1) | 1.0000 | 0.009(2) |

## III. Results and discussions

Figure 2 shows temperature-dependent magnetization along the *c*-axis and parallel to the *ab*-plane (in-plane and out-of-plane, respectively) in this material with an application of 0.1 T magnetic field (H). The M(T) curves show an anisotropic magnetic response for fields applied along different axes at temperatures below the transition temperature. Along the *c*-axis, a rapid upturn occurs around 168 K upon cooling, signifying a paramagnetic to ferromagnetic phase transition. The magnetization along the *c*-axis is about 13 times higher than that along the *ab*-plane. The M(T) measurement indicates a large uniaxial magnetic anisotropy in $Cr_2Te_3$. There was no significant difference between the zero-field-cooling (ZFC) and field-cooling (FC) data for fields applied along two directions signifying high crystal quality. To further explore the field-dependent magnetization along the easy and hard axis, we performed isothermal field-dependent magnetization at 2 K, as shown in Figure 3. The field-dependent magnetization at 2 K along the *c*-axis saturates at ~ 0.35 T field but does not saturate along the *ab*-plane even at 9 T field. The saturated magnetic moment per Cr atom at the 9 T field is 2.25 $\mu_B$/Cr for the H || *c*-axis and 2.14 $\mu_B$/Cr for the H || *ab*-plane. These results confirm that the *c*-axis is the magnetic easy axis of single-crystal $Cr_2Te_3$ (Figure 3).

To further explore magnetic anisotropy, we performed isothermal magnetization studies around its transition temperature (~170 K). Figures 4(a) and 4(b) show the isothermal magnetizations for fields up to 9 T applied along the *c*-axis and in the *ab*-plane, respectively, from 146 to 190 K, where data were taken every 2 K interval. Isothermal field-dependent magnetization measurements (M(H)) is performed using a Quantum Design Physical Property Measurement System (PPMS) 9 Tesla Dynacool model. There is a clear difference between field-dependent magnetization curves along the *c*-axis and in the *ab*-plane, particularly in the low-magnetic-field region. Below $T_C$, the magnetization along the *c*-axis tends to saturate for small fields, whereas it increases slowly at low fields along the *ab*-plane. This shows the presence of large magnetic anisotropy in $Cr_2Te_3$. The micromagnetic energy density can be expressed as $E_A = K_u \sin^2(\theta - \varphi)$ [41], where $K_u$ is the uniaxial magneto-crystalline parameter, $\theta$ is the direction of the preferred magnetization, $\phi$ is the direction in which the magnetization points. When $\theta - \phi = 90°$ (i.e., H || *ab*-plane), the magneto-crystalline anisotropy is maximum. The uniaxial anisotropy parameter $K_u$ is related to the saturation field $H_S$ and saturation magnetization $M_S$ by $2K_u/M_S = \mu_0 H_S$, where $\mu_0$ is the vacuum permeability. We estimated the $M_S$ by using a linear fit of M(H) curves above 6 T for the H || *c*-axis, which monotonically decreases with increasing temperature. We then determined $H_S$ as the intersection point of two linear fits: one being a fit to the saturated regime at a high field, and the other being a fit to the unsaturated linear regime at a low magnetic field. This procedure is explained in detail for the anisotropic magnetic entropy change studies of $VI_3$ [42] and $CrSbSe_3$ [36]. Figure 5 shows the temperature-dependent magnetic anisotropy, along with the temperature-dependent saturation magnetization and saturation magnetic field in the Figure 5 insets. The calculated $K_u$ decreases monotonically from 486.9185 kJ m$^{-3}$ at 146 K to 150.6002 kJ m$^{-3}$ near $T_C$. The magnetic anisotropy constants are independent of temperature. But the observed decrease in $K_u$ with increasing temperature arises solely from many local spin clusters fluctuating randomly around the macroscopic magnetization vector and activated by nonzero thermal energy [43,44].

To understand magnetism in the single-crystal Cr$_2$Te$_3$ system, we estimate changes in magnetic entropy along different axes in this material. The change of magnetic entropy $\Delta S_M$ (T, H) is defined as the following [31,45]:

$$\Delta S_M(T,H) = \int_0^H \left[\frac{\partial S(T,H)}{\partial H}\right] dH \quad (1a)$$

Using the Maxwell's relation [$\partial S(T, H)/\partial H = \partial M(T, H)/\partial T$], this can be expressed as following:

$$\Delta S_M(T,H) = \int_0^H \left[\frac{\partial M(T,H)}{\partial T}\right] dH \quad (1b)$$

In the case of isothermal magnetization measured at small discrete magnetic fields and temperature intervals, $\Delta S_M$(T, H) is the isothermal change of entropy in the interval of the magnetic field from 0 to H and can be approximated as

$$\Delta S_M(T,H) = \frac{\int_0^H M(T_{i+1},H) - \int_0^H M(T_i,H)}{T_{i+1} - T_i} \quad (2)$$

The calculated $-\Delta S_M$ values as a function of temperature in various fields up to 9 T applied along the *c*-axis and *ab*-plane are shown in Figures 6(a) and 6(b), respectively. All these $-\Delta S_M$(T, H) curves show a pronounced peak around the phase transition temperature, and the peak broadens asymmetrically on both sides with increasing field. Here, it is clear from these two figures that the $\Delta S_M$ curves along the *ab*-plane are much more asymmetric in comparison to $\Delta S_M$ curves along the *c*-axis. Also, $-\Delta S_M$ reaches a maximum of ~4.5 J kg$^{-1}$ K$^{-1}$ along the *c*-axis and ~3.36 J kg$^{-1}$ K$^{-1}$ along the *ab*-plane. The rotational magnetic entropy changes $\Delta S^R_M$ induced by rotating the field from the *c*-axis to the *ab*-plane can be expressed as:

$$\Delta S^R_M = \Delta S_M(T, H_c) - \Delta S_M(T, H_{ab}) \quad (3)$$

Figure 6(c) shows temperature-dependent rotational magnetic entropy change for different fields. The anisotropy is gradually suppressed at a higher field, as seen in Figure 6(c), and interestingly, it splits into many peaks (in scattered pattern) on both sides of the $T_C$ for field above 3 T for single-crystal Cr$_2$Te$_3$. Up to a low magnetic field (3 T), the peak near $T_C$ is clearly observed, but above 3 T, there is a discrepancy in the pattern of the curves, which is an indication of decreasing anisotropy due to increasing magnetic field.

To further explore the nature of its magnetic phase transition, we analyzed the $\Delta S_M$ data following the proposed universal scaling analysis [46]. It is constructed by normalizing all $-\Delta S_M$ curves against the respective maximum $-\Delta S_M^{max}$, namely, $\Delta S_M/\Delta S_M^{max}$, and then by rescaling the temperature θ below and above $T_C$, as defined in the following equations:

$$\theta_- = (T_{peak} - T)/(T_{r1} - T_{peak}), \quad T < T_{peak}, \quad (4)$$

$$\theta_+ = (T - T_{peak})/(T_{r2} - T_{peak}), \quad T > T_{peak}, \quad (5)$$

where $T_{r1}$ and $T_{r2}$ are the temperatures of the two reference points selected as those corresponding to $\Delta S_M(T_{r1}, T_{r2}) = \Delta S_M^{max}/2$. As shown in Figures 7(a)-(b), all $-\Delta S_M(T, H)$ curves in various fields collapse into a single curve, implying a second-order magnetic phase transition for single-crystal $Cr_2Te_3$.

For a material undergoing a second-order magnetic phase transition [47], the various parameters of $\Delta S_M(T)$ curves obey different field-dependent power laws, as following [48,49]:

$$|\Delta S_M^{max}| = a\, H^n, \quad (6)$$

$$\delta T_{FWHM} = m\, H^b, \quad (7)$$

$$RCP = d\, H^c, \quad (8)$$

Here $\Delta S_M^{max}$ is the maximum value of $\Delta S_M$, $\delta T_{FWHM}$ is the full width at half maximum of $\Delta S_M(T, H)$ curve and RCP is the relative cooling power. In the case of magnetic refrigerators, RCP determines the cooling efficiency of the refrigerant and corresponds to the amount of heat that can be transferred between the cold and hot parts of the refrigerator in an ideal thermodynamic cycle [30,46]. This process plays a vital role in magnetic refrigerators. RCP is expressed as:

$$RCP = \Delta S_M^{max} \times \delta T_{FWHM} \quad (9)$$

Therefore, $|\Delta S_M^{max}|$, RCP, and $\delta T_{FWHM}$ are related to the applied magnetic field with power law relation, as expressed in equations (6), (7), and (8). Here exponent $c$ is related to the critical exponent δ, and the other exponents $n$ and $b$ are related to the other critical exponents β and γ, as following [46]:

$$c = 1 + 1/\delta, \quad (10a)$$

$$b = 1/\Delta = 1/\beta\delta, \quad (10b)$$

$$n(T_C) = 1 + \left[\frac{\beta-1}{\beta+\gamma}\right] = 1 + 1/\delta\left(1 - \frac{1}{\beta}\right). \quad (10c)$$

Here, β, γ, and δ are called critical exponents [50]. Therefore, by analyzing the magneto-caloric effect (MCE) we will be able to determine these critical exponents. The critical exponents obtained from the MCE analysis are more reliable because it is a direct method to determine these critical exponents [51,52]. These estimated exponents can be verified by Widom's law: $\delta = 1+\gamma/\beta$ [53]. Figures 8(a) and 8(b) represent the field dependence of $-\Delta S_M^{max}$ and RCP for the H ∥ c-axis and H ∥ ab-plane, respectively. The calculated values of RCP are approximately 296.6969 and 183.8444 J kg$^{-1}$ for the H || c-axis and H || ab-plane, respectively, with a magnetic field change of 9 T. Fitting of $-\Delta S_M^{max}$ gives n = 0.6507±0.0070 and 1.0235±0.0050 for the H || c-axis and H || ab-plane, respectively [Figures 8(a)-(b)]. Among these values, the n value (for H || c-axis) was quite close to that of n = 0.6667 for the mean-field model (β = 0.5, γ = 1). Fitting of the RCP generates c = 1.1257±0.0010 and 1.5897±0.0020 [Figures 8(a)-(b)], which give δ = 7.9491 and 1.6957 for the H || c-axis and H || ab-plane, respectively.

Similarly, Figures 9(a)-(b) show the magnetic field dependent $\delta T_{FWHM}$ for H || c-axis and H || ab-plane, respectively, and estimated values of $\delta T_{FWHM}$ are 65.9192 K and 54.6863 K respectively, for H || c-axis and H || ab-plane with a magnetic field change of 9 T. The curves fit well with equation (7) and fittings of the $\delta T_{FWHM}$ curves give b = 0.5295±0.0040 and 0.6486±0.0067 for H || c-axis and H || ab-plane, respectively [Figures 9(a)-(b)]. Hence using the equation (10b) we determined Δ = 1.8885 and 1.5418 for H || c-axis and H || ab-plane, respectively [Figures 9(a)-(b)]. After solving all these equations, we got β = 0.2435, γ = 1.6509, and δ = 7.9491 respectively for H || c-axis whereas we got β = 0.9092, γ = 0.6325 and δ = 1.6957 for H || ab-plane. The estimated critical exponent values for the H || ab-plane are different from those estimated values for the H || c-axis. We will verify their reliability later by universal scaling analysis. Similar differences in estimated critical exponent values for two axes have been reported earlier for VI$_3$ [54]. Table II shows our estimated critical exponents with other reported and numerically estimated values.

For a material undergoing a second-order magnetic phase transition, the maximum magnetic entropy change follows a power law dependence with applied magnetic field: $-\Delta S_M^{max} = aH^n$, where a is a constant and the exponent n is related to the magnetic order. This can be expressed in another way $n(T, H) = d \ln|\Delta S_M|/d \ln(H)$. Using this expression, all n(T) curves for each magnetic field can be estimated. Figure 10 shows the temperature dependence of n(T) for various fields. All n(T) curves follow a pattern. At low temperatures, well below T$_C$, n(T) is about 1. On the other side, well above T$_C$, n is close to 2 because of the Curie-Weiss law. At T = T$_C$, n(T) has a minimum. It could be found that with a field change of 9 T, the n values are 1.09 and 1.81 far below and above T$_C$, respectively, consistent with the universal law of the n [49]. With the reduction of the magnetic field, the value of n is nearly unchanged at T$_C$ and higher temperatures, however, it shows deviation at lower temperatures, which might be due to the magnetic anisotropy effect.

For a second-order magnetic phase transition, the $\Delta S_M(H)$ vs. T curves under different external fields can be rescaled by the scaled equation of state: $H/M^\delta = f(\varepsilon/M^{1/\beta})$, where δ and β are critical exponents [32]. Here the reduced temperature $\varepsilon = (T - T_C)/T_C$. The $\Delta S_M(T, H)$ can be rewritten in another form:

$$\Delta s_M(T, H) = H^{(1-\alpha)/\Delta} g\left(\varepsilon/H^{1/\Delta}\right) \quad (11)$$

Figures 11(a)-(b) show $-\Delta S_M/H^{(1-\alpha)/\Delta}$ vs. $\varepsilon/H^{1/\Delta}$ for H || c-axis and H || ab-plane, respectively, which show that all rescaled curves under different fields and temperatures collapse into a single universal curve. The exact collapse and overlapping of these curves confirm the reliability of the critical exponents estimated by the magnetic entropy change method.

The estimated critical exponent values are listed in Table II along with other theoretical and experimental values. Taroni *et al*. have shown that the critical exponent β for 2D magnets is within a range of 0.1 ≤ β ≤ 0.25 [55]. Our estimated β value lying in this range indicates that $Cr_2Te_3$ might behave like a 2D magnetic material. We also find that β is very close to that of the tri-critical mean-field model, while γ value approaches the 2D-Ising model value. All these imply that magnetic interactions in $Cr_2Te_3$ cannot be simply described as 2D-Ising type interactions, but multiple interactions are present in this system.

Based on the critical exponents obtained above, the range of magnetic exchange interaction in $Cr_2Te_3$ can be estimated. For a homogeneous magnet, the universality class of the magnetic phase transition depends on the exchange distance J(r). Fisher *et al*. theoretically treated this kind of magnetic ordering as an attractive interaction of spins, in which the renormalization group theory analysis suggests that the interaction decays with distance *r* as $J(r) \sim 1/r^{(d+\sigma)}$, where *d* is the spatial dimensionality, and σ is a positive constant [56]. Moreover, the critical exponents γ can be expressed as following:

$$\gamma = 1 + \frac{4}{d}\left(\frac{n+2}{n+8}\right)\Delta\sigma + \frac{8(n+2)(n-4)}{d^2(n+8)^2} \times \left[1 + \frac{2G(d/2)(7n+20)}{(n-4)(n+8)}\right]\Delta\sigma^2 \quad (12)$$

Here $G\left(\frac{d}{2}\right) = 3 - \frac{1}{4}\left(\frac{d}{2}\right)^2$, Δσ = σ – d/2, and *n* is the spin dimensionality. Using *d* = 2 in equation (12), our estimated σ values are 1.5396 and 1.4601 for spin dimensionality *n* = 1 and 2, respectively. The correlation length critical exponent υ = γ/σ gives us the critical exponents α = 2 − υd ≈ −0.1446 and −0.2613 for *n* = 1 and 2, respectively. Moreover, applying the Rushbrooke inequality relation [50], we evaluate α = -0.1379 from our estimated β and γ value. This α = -0.1379 value is very close to the above estimated value of α = -0.1446 for *n* = 1.

For the spatial dimension *d* = 3, computed σ values using equation (12) are 2.1433, 2.2186, and 2.3469 for *n* = 1, 2, and 3, respectively. After putting α value in above-mentioned procedure we calculated α = -0.3108, -0.2323, and -0.1103 for *n* = 1, 2, and 3, respectively. The calculated α values are not good match with the estimated α = -0.1379 from Rushbrooke inequality relation [50] using the critical exponents estimated from the magnetocaloric effect (MCE). This is confirmation that critical exponents estimated for $Cr_2Te_3$ from MCE are close to those for the 2D system with spin dimensionality *n* = 1. Moreover, σ < 2 shows the presence of a long-range magnetic ordering in this material. As for a 2D system, long range magnetic ordering exists due to the presence of strong magneto-crystalline anisotropy which can reduce thermal fluctuation [57].

$Cr_2Te_3$ is not a layered van der Walls (vdW) material like $CrI_3$ [1], $Cr_2Ge_2Te_6$ [1], etc. For $Cr_2Te_3$, magnetism originates from the magnetic ordering of Cr moments. The neutron diffraction revealed the ferromagnetic ordering of magnetic moments along the *c*-axis and displayed a reduction of the unit cell volume as the temperature decreased [27,28]. In this material, there are covalently bonded Cr atoms between layers of Cr-Te atoms, as shown in Figure 1(b) [22,58]. This may explain the two-dimensional nature of magnetism of $Cr_2Te_3$ as seen in our analysis.

Temperature dependence of heat capacity $C_p$ for single crystal $Cr_2Te_3$ measured in constant magnetic fields (up to 9 T) applied along the *c*-axis is shown in Figure 12(a). A sharp λ-type peak is observed at ~171 K in zero magnetic field $C_p$ which indicates a paramagnetic to ferromagnetic phase transition. With an increase of the magnetic field, the peak's height is reduced, and width is broadened. The peak also shifts toward the higher temperature with an increasing magnetic field. The estimated heat capacity change $\Delta C_p = C_p(T, H) - C_p(T, 0)$ as a function of temperature in various fields is plotted in Figure 12(b). It should be noted that $\Delta C_p < 0$ for $T < T_C$ and $\Delta C_p > 0$ for $T > T_C$ and it changes rapidly from negative to positive value at $T_C$. This rapid upturn shown in Figure 12(b) confirms a paramagnetic to ferromagnetic phase transition in this material. The total entropy S(T, H) can be calculated from heat capacity data by the following relation:

$$S(T,H) = \int_0^T \frac{c_P(T,H)}{T} dT \qquad (12)$$

Here it is assumed that the electronic and lattice contributions are not field-dependent [59]. In an adiabatic magnetization process, the magnetic entropy change $\Delta S_M$ should be $\Delta S_M(T, H) = \Delta S_M(T, H) - \Delta S_M(T, 0)$. Next, we are going to estimate the adiabatic temperature change $\Delta T_{ad}$ for various temperatures around the magnetic phase transition. The adiabatic temperature change is defined as the temperature change of the material when adiabatically magnetized/demagnetized. The adiabatic temperature change caused by the field change can be indirectly determined, $\Delta T_{ad}(S, H) = T(S, H) - T(S, 0)$, where T(S, H) and T(S, 0) are the temperatures in the field $H \neq 0$ and $H = 0$, respectively, at a constant entropy [32]. Figures 12(c)-(d) show $\Delta T_{ad}$ and $\Delta S_M$, respectively, estimated from heat capacity data as a function of temperature for various applied fields. All these curves reach maximum near the Curie temperature and increase with the increasing of magnetic field. In figure 12(c), with the enhancement of magnetic field, $\Delta T_{ad}$ curves broaden and the peak is shifted to the higher temperature just like temperature dependent specific heat graph. It is worth noting that peak of the $\Delta T_{ad}$ is shifted from 171 K at 1 T to 175 K at 9 T magnetic field. The maxima of $-\Delta S_M$ and $\Delta T_{ad}$ reach the values of 5.25 J kg$^{-1}$ K$^{-1}$ (whereas 4.5 J kg$^{-1}$ K$^{-1}$ from isothermal magnetization) and 3.31 K, respectively, for 9 T magnetic field.

## IV.   Conclusion

In summary, we have systematically studied the anisotropic magnetocaloric effects (MCE) of single crystal $Cr_2Te_3$. We have grown these $Cr_2Te_3$ samples by CVT and characterized by x-ray diffraction. The second-order nature of the paramagnetic-to-ferromagnetic phase transition near $T_C = 171$ K has been verified by the scaling analysis of magnetic entropy change. The critical exponents β, γ, and δ are estimated by analyzing the magnetocaloric effect, and the scaling analysis of magnetic entropy change confirms the accuracy of our estimated critical exponents. A large magneto-crystalline anisotropy constant $K_u$ is estimated to be 486.9158 kJ m$^{-3}$ at 146 K. Using the heat capacity measurements, we estimate adiabatic temperature $\Delta T_{ad} = 3.53$ K for 9 T magnetic field. Our analysis shows that $Cr_2Te_3$ behaves like a 2D Ising system with a long-range magnetic ordering.

**Acknowledgements**

This work is supported by the National Science Foundation Awards No. DMR-2018579 and No. DMR-2302436. This work made use of the synthesis facility of the Platform for the Accelerated Realization, Analysis, and Discovery of Interface Materials (PARADIM), which is supported by the National Science Foundation under Cooperative Agreement No. DMR-2039380. This research used beamline 28-ID-1 of the National Synchrotron Light Source-II, a U.S. Department of Energy (DOE) Office of Science User Facility operated for the DOE Office of Science by Brookhaven National Laboratory under Contract No. DE-SC0012704.

*Address of correspondence: samaresh.guchhait@Howard.edu

Table II: Estimated critical exponent values of $Cr_2Te_3$, including other experimental and theoretical critical exponent values.

| Composition | techniques | β | γ | δ |
|---|---|---|---|---|
| $Cr_2Te_3$ (this work) | Magnetocaloric Effect (MCE) | 0.2435 | 1.6509 | 7.9491 |
| Theory [60] | Mean-field model | 0.5 | 1 | 3 |
| Theory [61] | 3D Heisenberg model | 0.365 | 1.386 | 4.82 |
| Theory [61] | 3D XY model | 0.345 | 1.361 | 4.81 |
| Theory [61] | 3D Ising model | 0.325 | 1.24 | 4.80 |
| Theory [61] | Tri-critical mean field model | 0.25 | 1 | 5 |
| Theory [62] | 2D Ising model | 0.125 | 1.75 | 15 |
| $Cr_5Te_6$ [16] | Kouvel-Fisher plot | 0.406 | 1.199 | 3.99 |
| $Cr_4Te_5$ [20] | Kouvel-Fisher plot | 0.387 | 1.287 | 4.32 |

| Cr$_{0.62}$Te [17] | Kouvel-Fisher plot | 0.315 | 1.81 | 6.75 |
| Cr$_5$Te$_8$ [18] | Kouvel-Fisher plot | 0.321 | 1.27 | 4.9 |

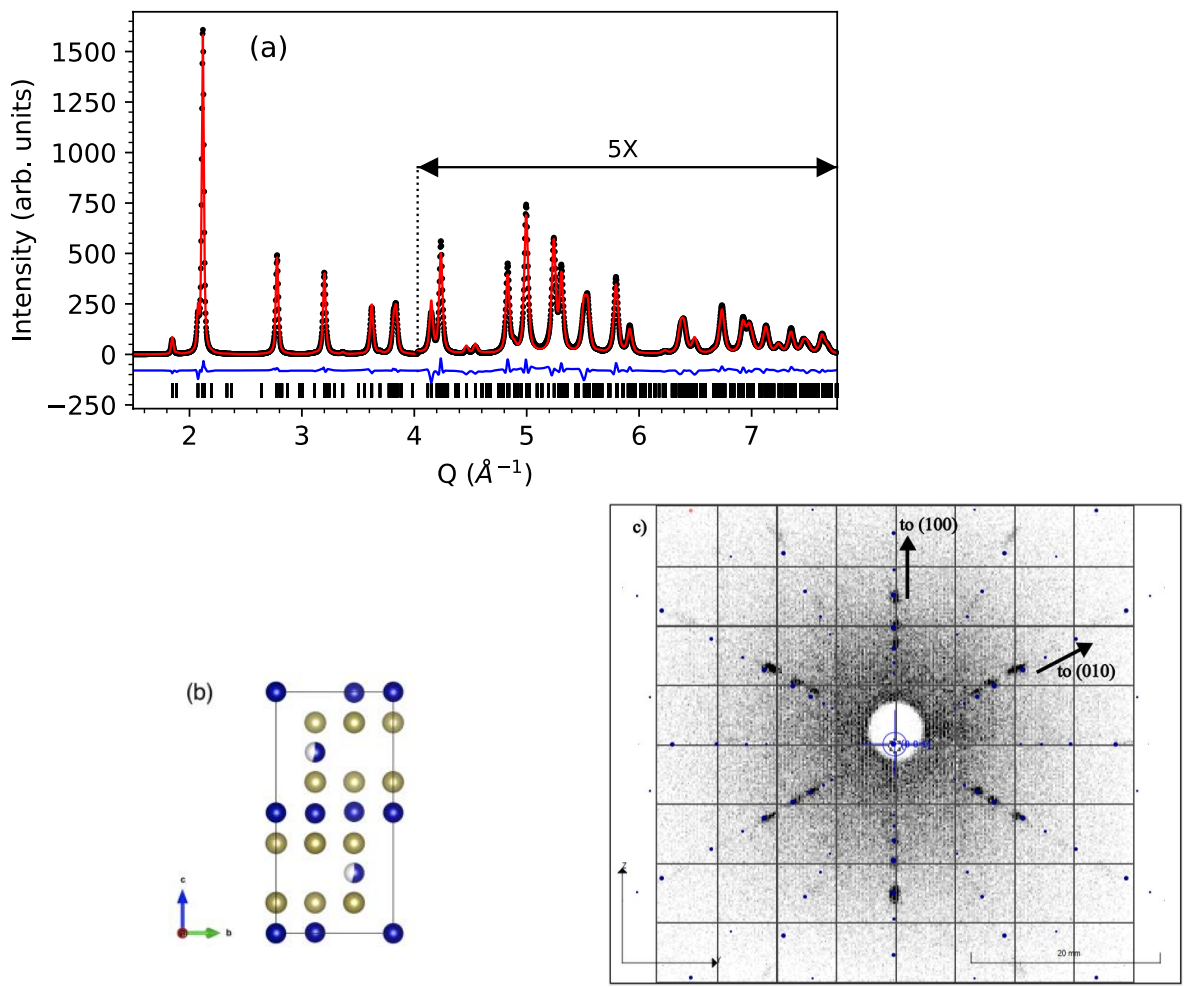

**Figure 1:** (a) Rietveld fit to background subtracted powder diffraction data (Goodness-of-fit, Rwp = 5.3 %). Black dots and the red line represent measured and calculated intensities, respectively. A residue plot (blue), and calculated Bragg reflection tick marks are shown below. The observed and calculated intensities, and the fit residue above Q ~ 4 Å$^{-1}$ are multiplied by 5 for clarity. (b) View of the Refined crystallographic unit cell along a crystallographic axis. Gold, Blue, and half blue half white balls represent Te, Cr, and Cr deficient sites, respectively. (c) Laue diffraction image of the single crystal $Cr_2Te_3$ sample. The sample is oriented along the crystallographic (001) axis. Arrows show the direction to the other crystallographic axes, as simulated by the QLaue software.

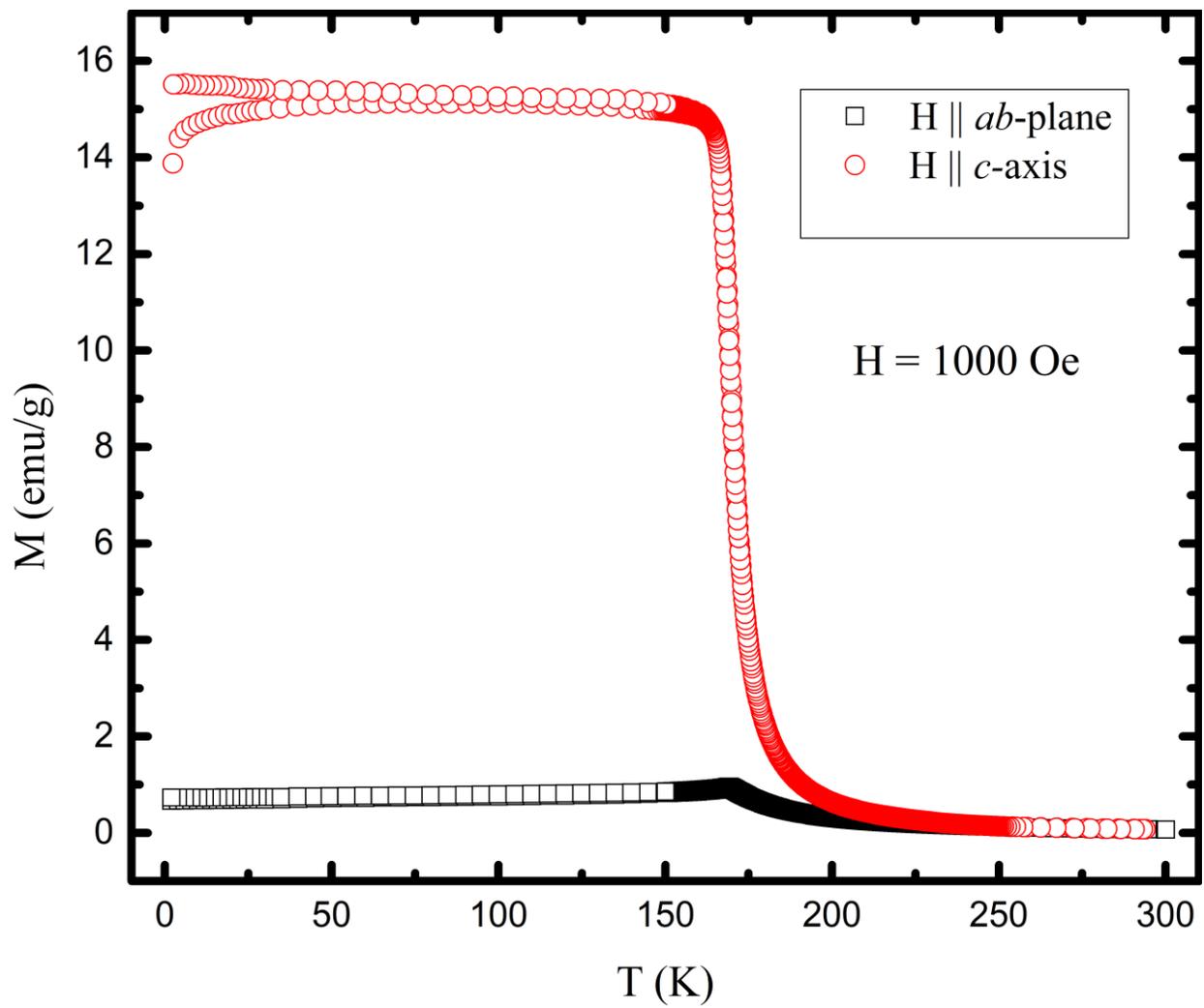

Figure 2: Temperature dependent zero-field-cooled (ZFC) and field-cooled (FC) magnetization for 0.1 T magnetic field applied along the *c*-axis (in-plane) and *ab*-plane (out-of-plane).

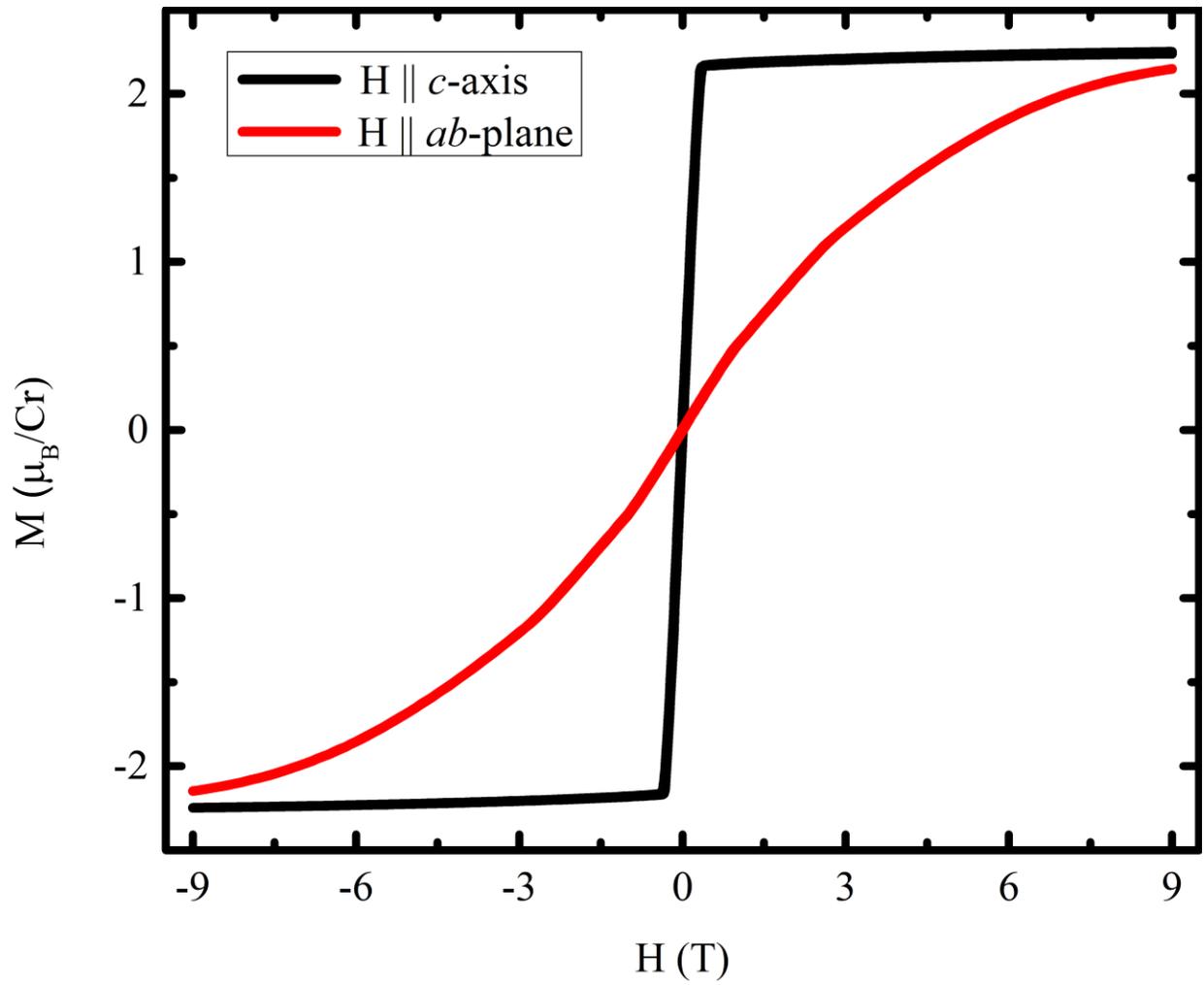

Figure 3: Field dependence of magnetization M(H) at 2 K for field applied along the *c*-axis and *ab*-plane of a single crystal sample.

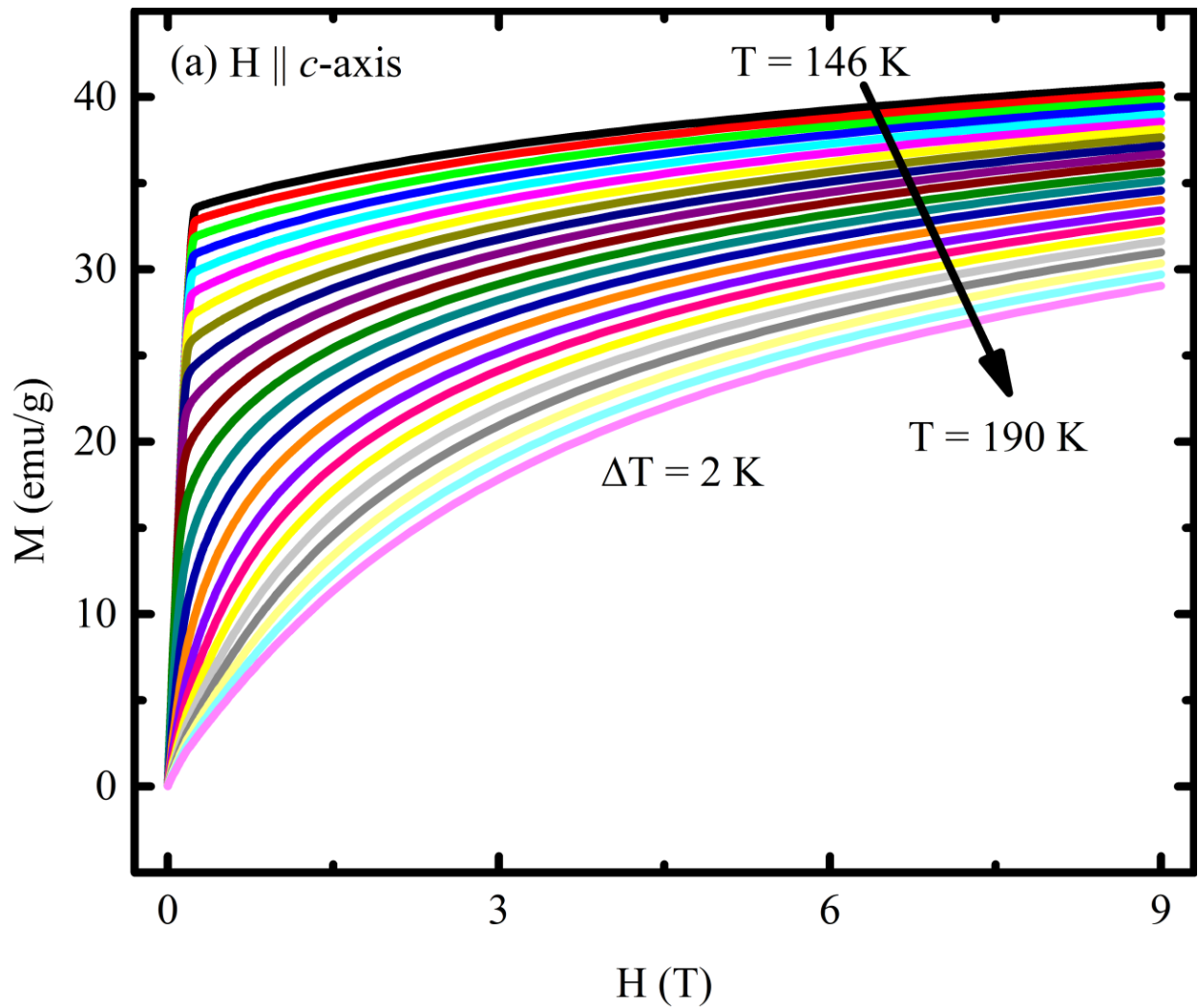

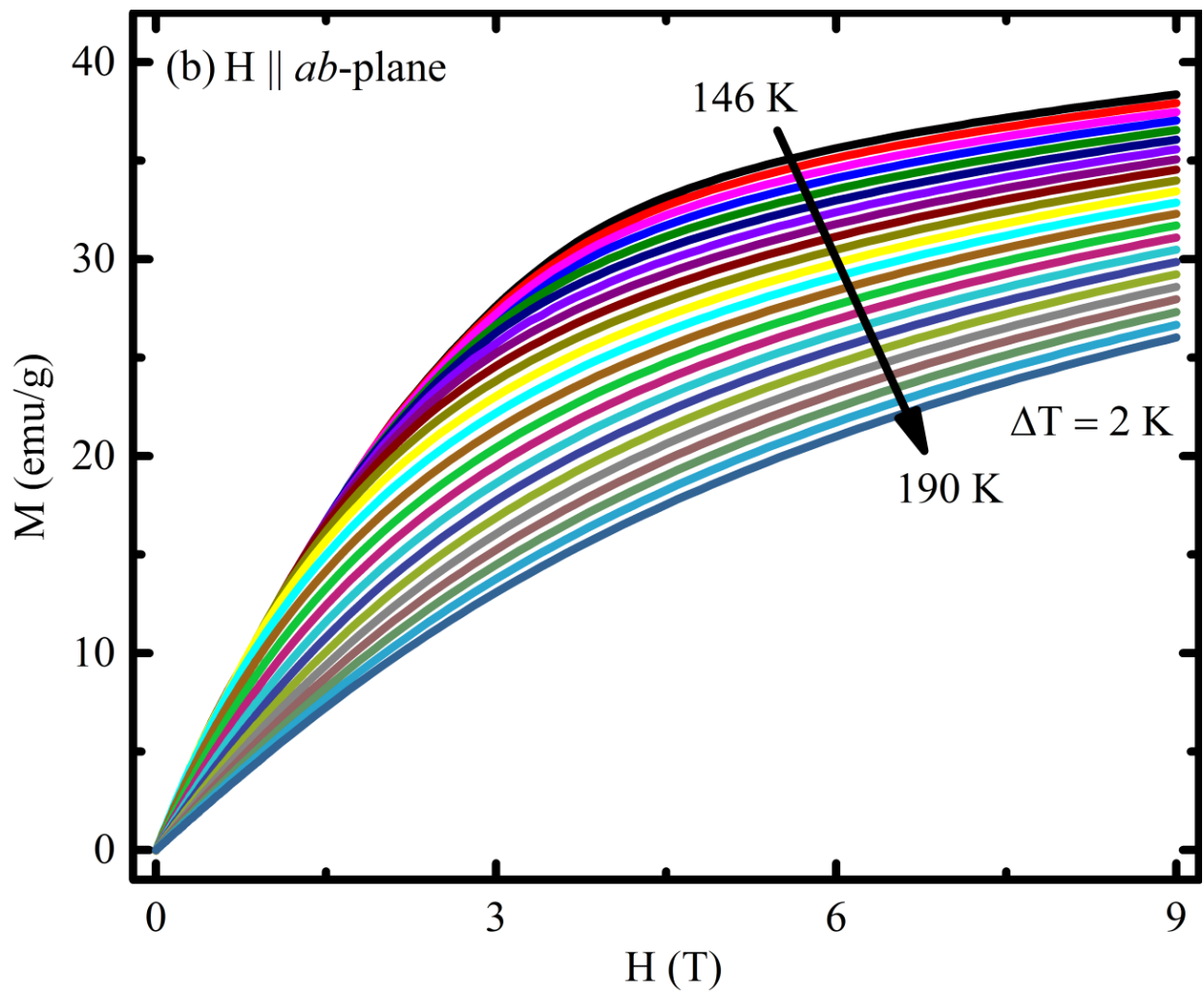

Figure 4: Isothermal magnetizations M(H) along the (a) *c*-axis and (b) *ab*-plane around the paramagnetic to ferromagnetic transition from 146 K to 190 K.

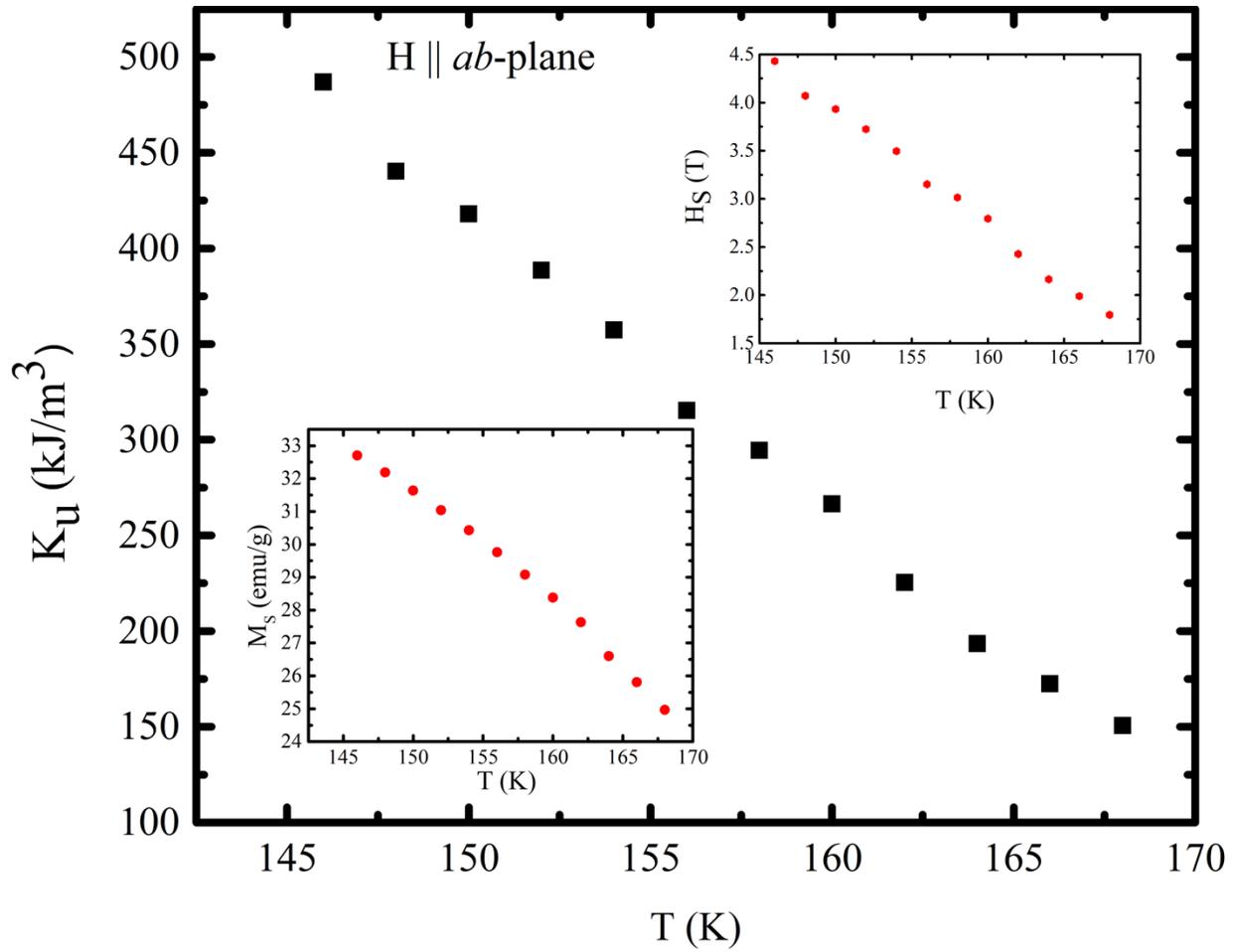

Figure 5: Temperature dependence of the calculated anisotropy constant $K_u$, the estimated saturation field $H_S$ (inset), and the saturation magnetization $M_S$ (inset) below transition temperature.

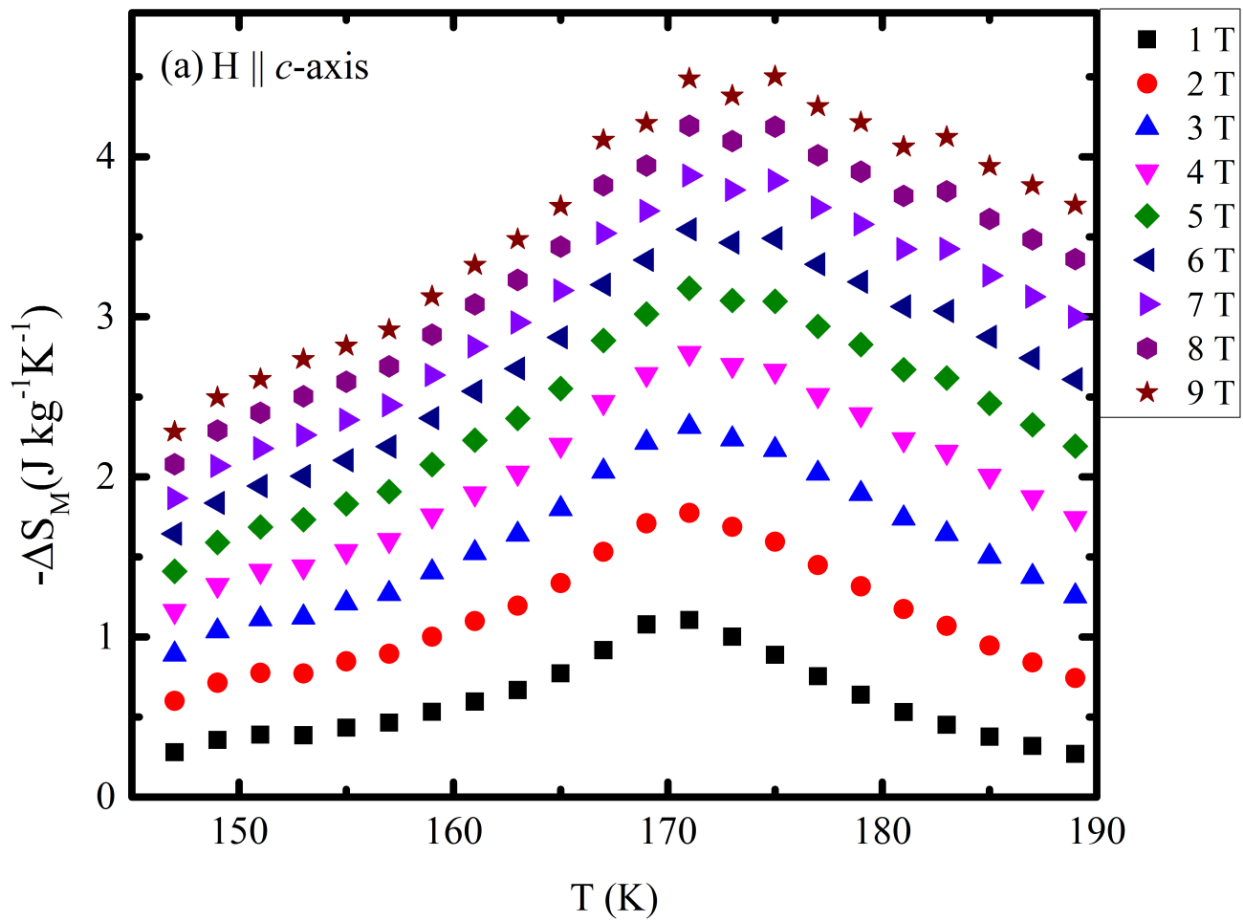

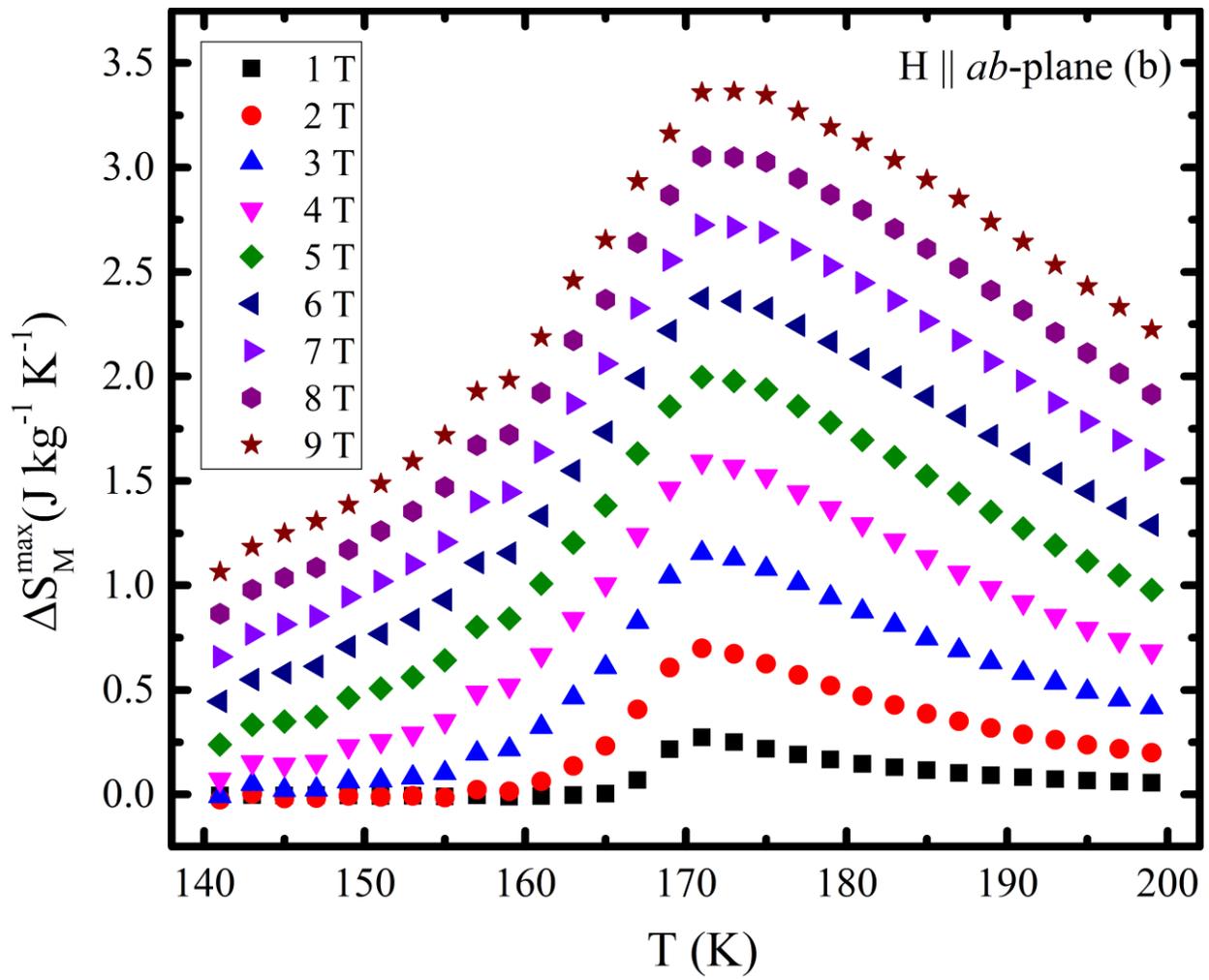

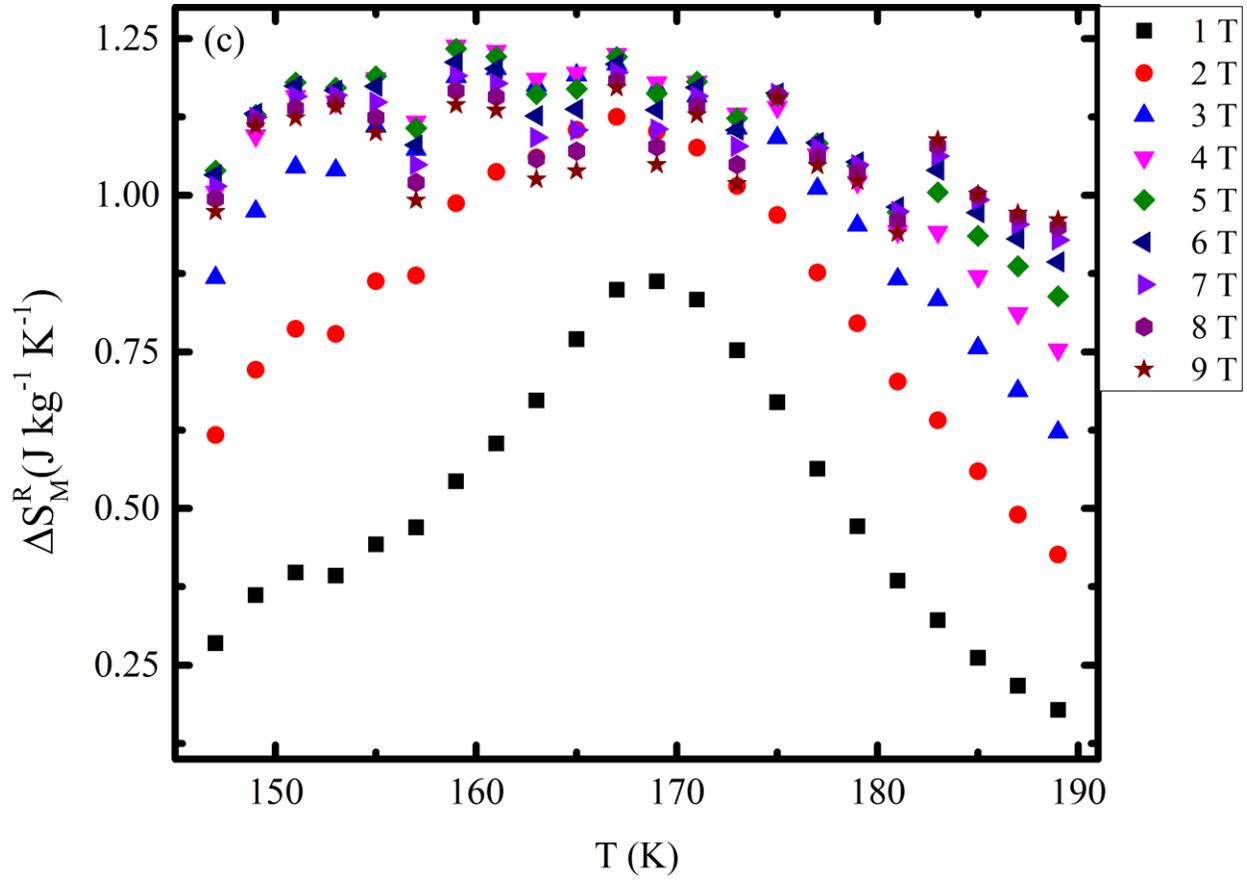

Figure 6: Temperature dependence change of magnetic entropy with increasing magnetic field along the (a) *c*-axis and (b) the *ab*-plane near the paramagnetic to ferromagnetic phase transition. (c) Temperature dependence of magnetic entropy change $-\Delta S^R_M$ obtained by rotating from the *ab*-plane to the *c*-axis in various fields.

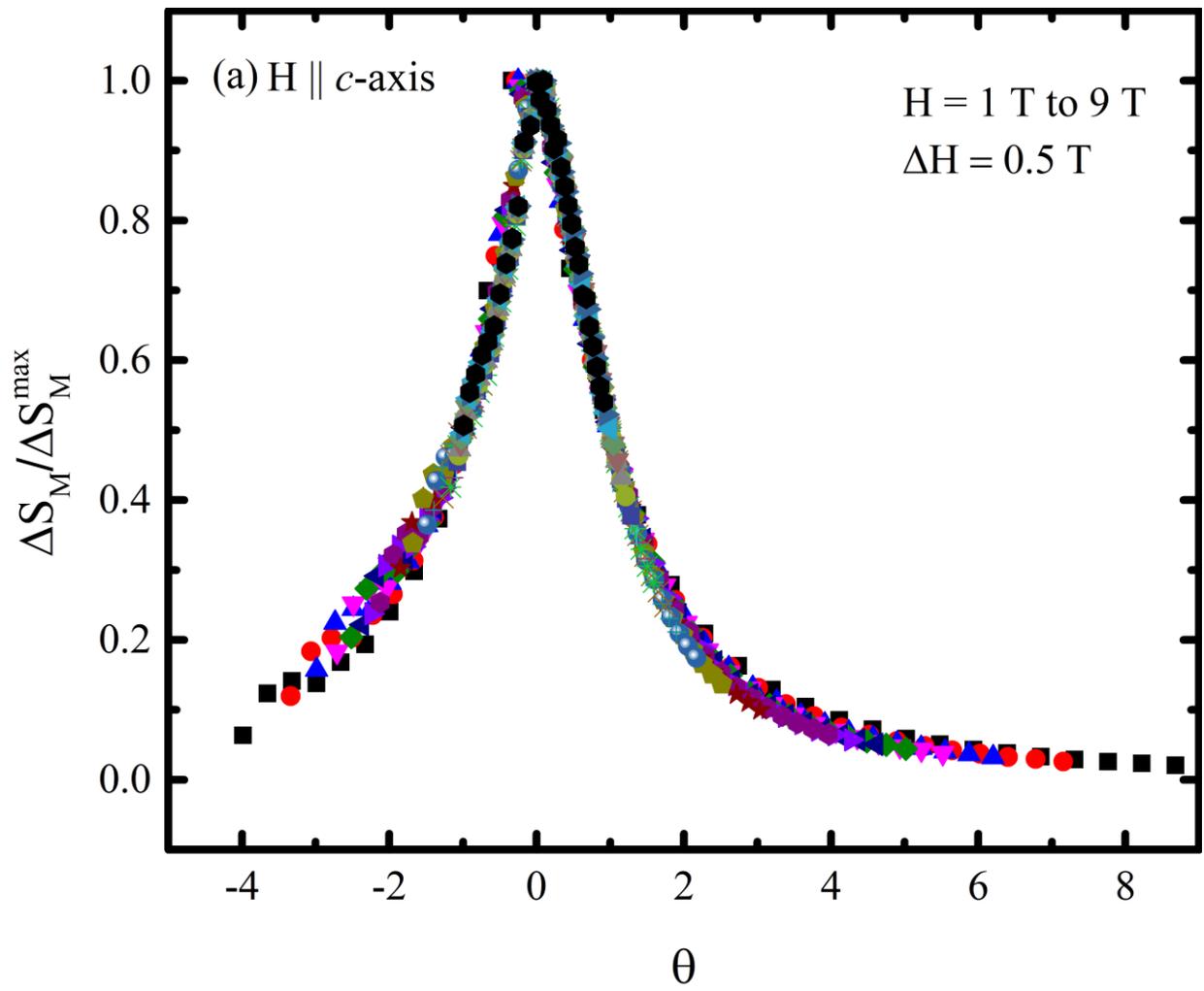

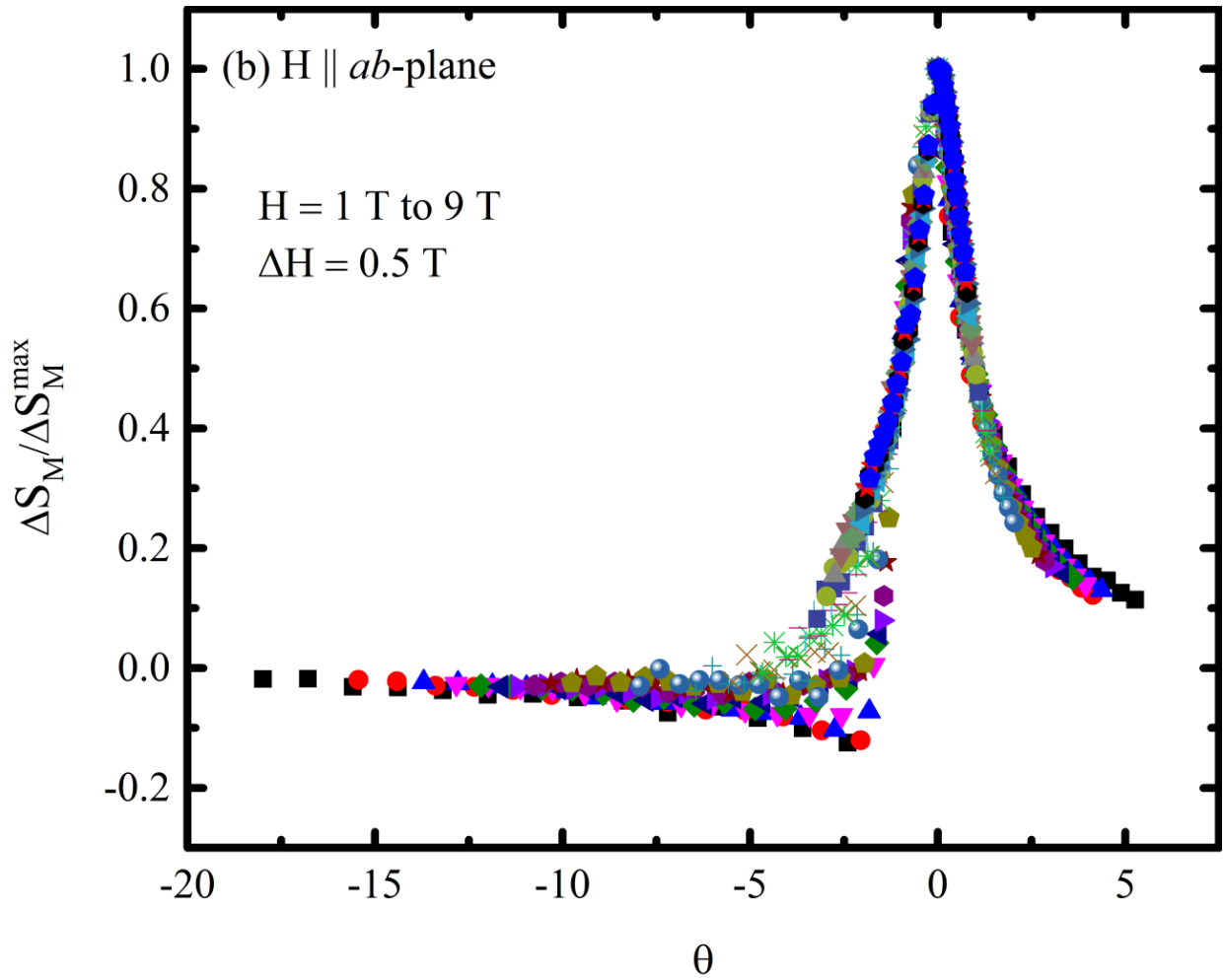

Figure 7: The normalized magnetic entropy ($\Delta S_M/\Delta S_M^{max}$) as a function of rescaled temperature $\theta$ along the (a) *c*-axis and (b) *ab*-plane.

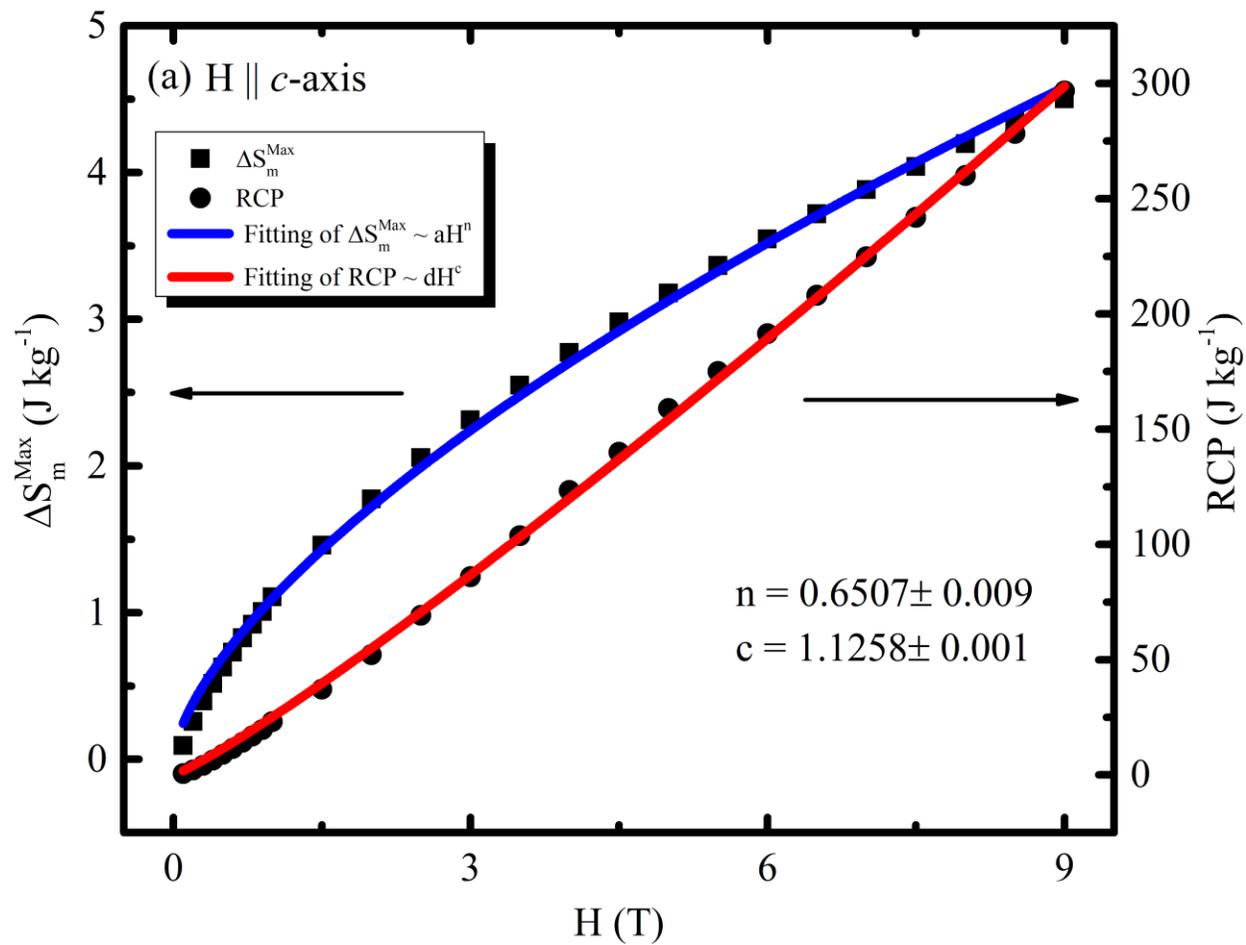

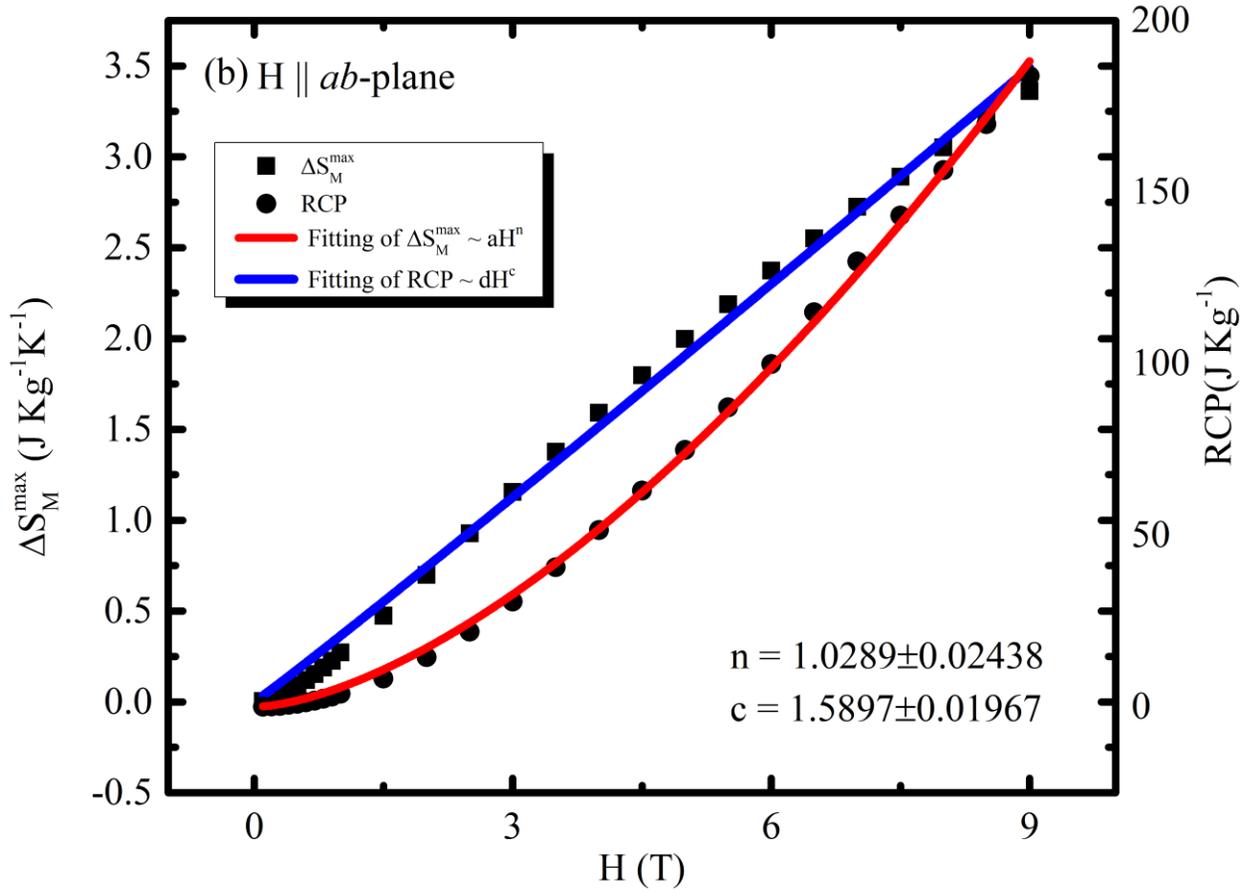

Figure 8: Magnetic field dependence of the maximum magnetic entropy change $-\Delta S^{max}_M$ and the relative cooling power (RCP) with power law fitting in blue and red solid lines, respectively with applied field along the (a) *c*-axis and (b) *ab*-plane.

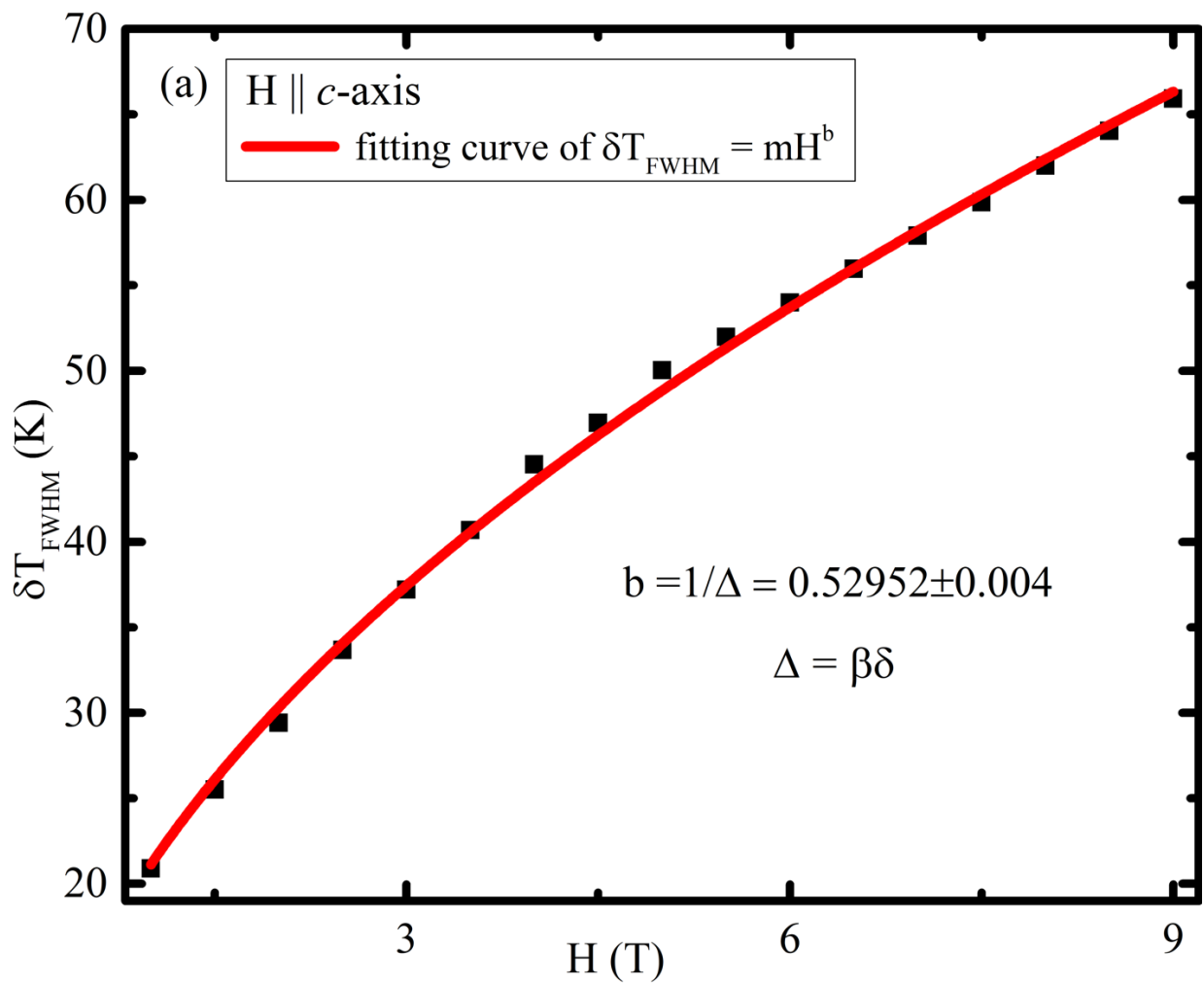

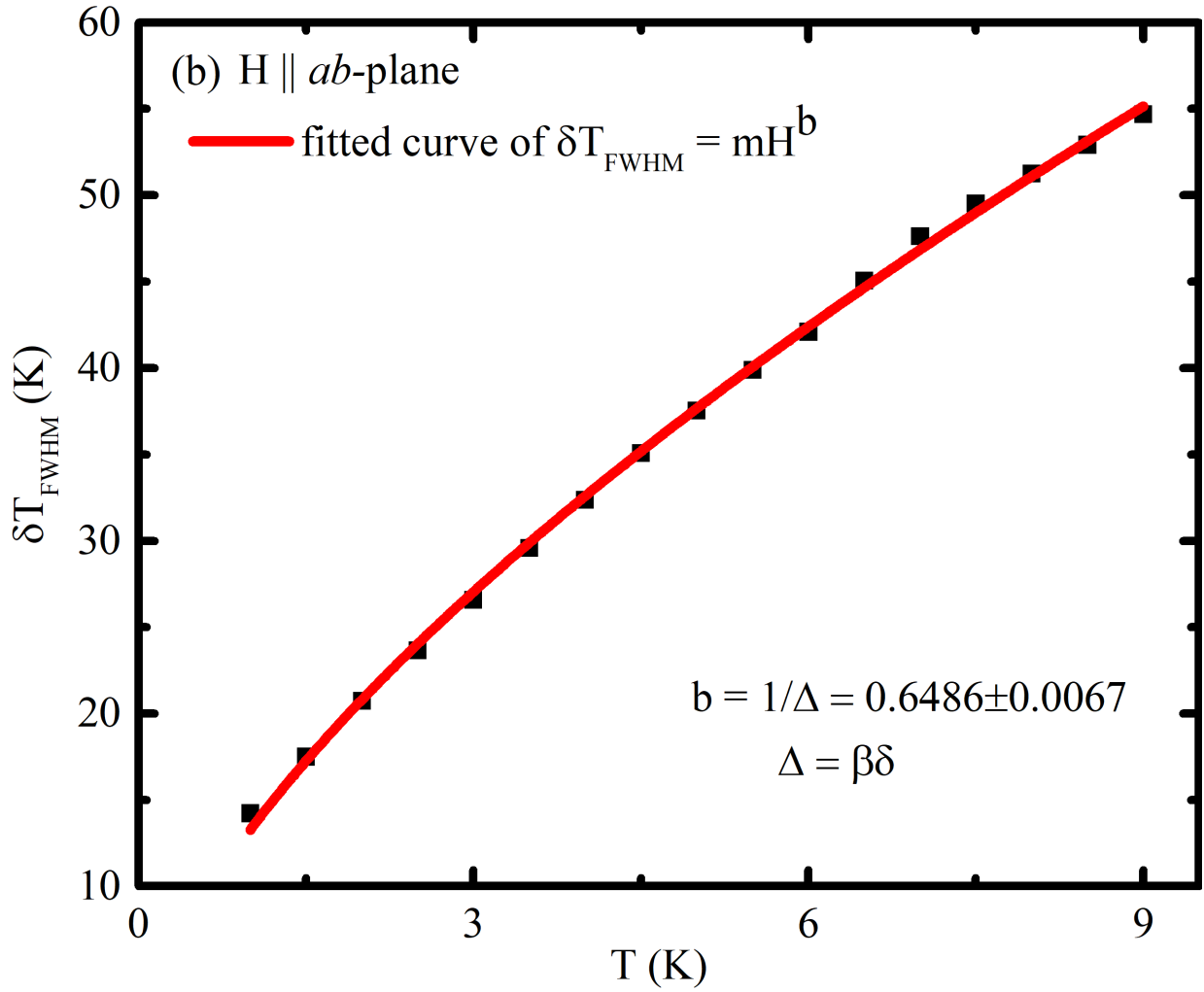

Figure 9: Magnetic field dependence of calculated δT$_{FWHM}$ with fitting curve when magnetic field applied along the (a) *c*-axis and (b) *ab*-plane.

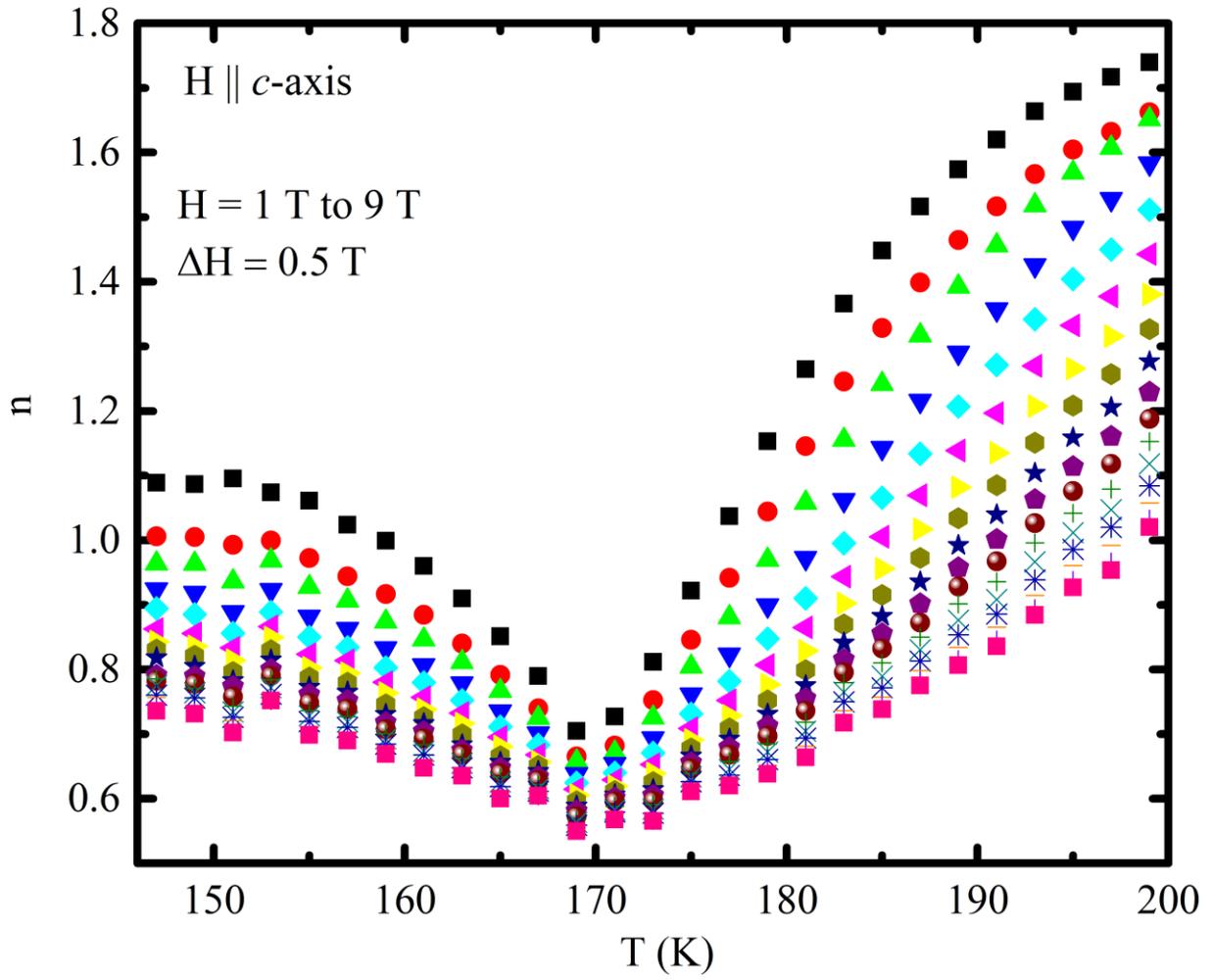

*Fig 10. Temperature dependence of n for various magnetic fields applied along the c-axis.*

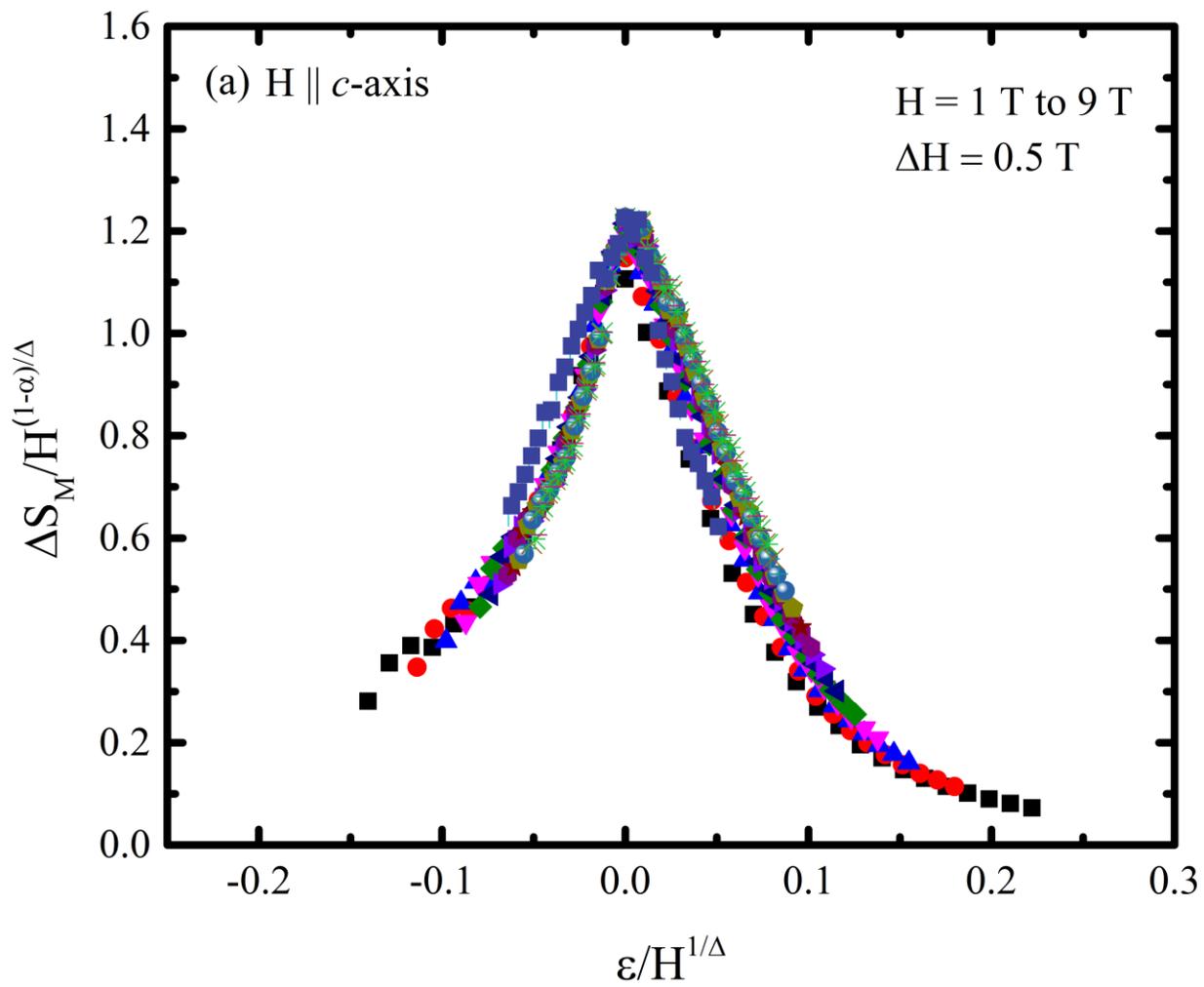

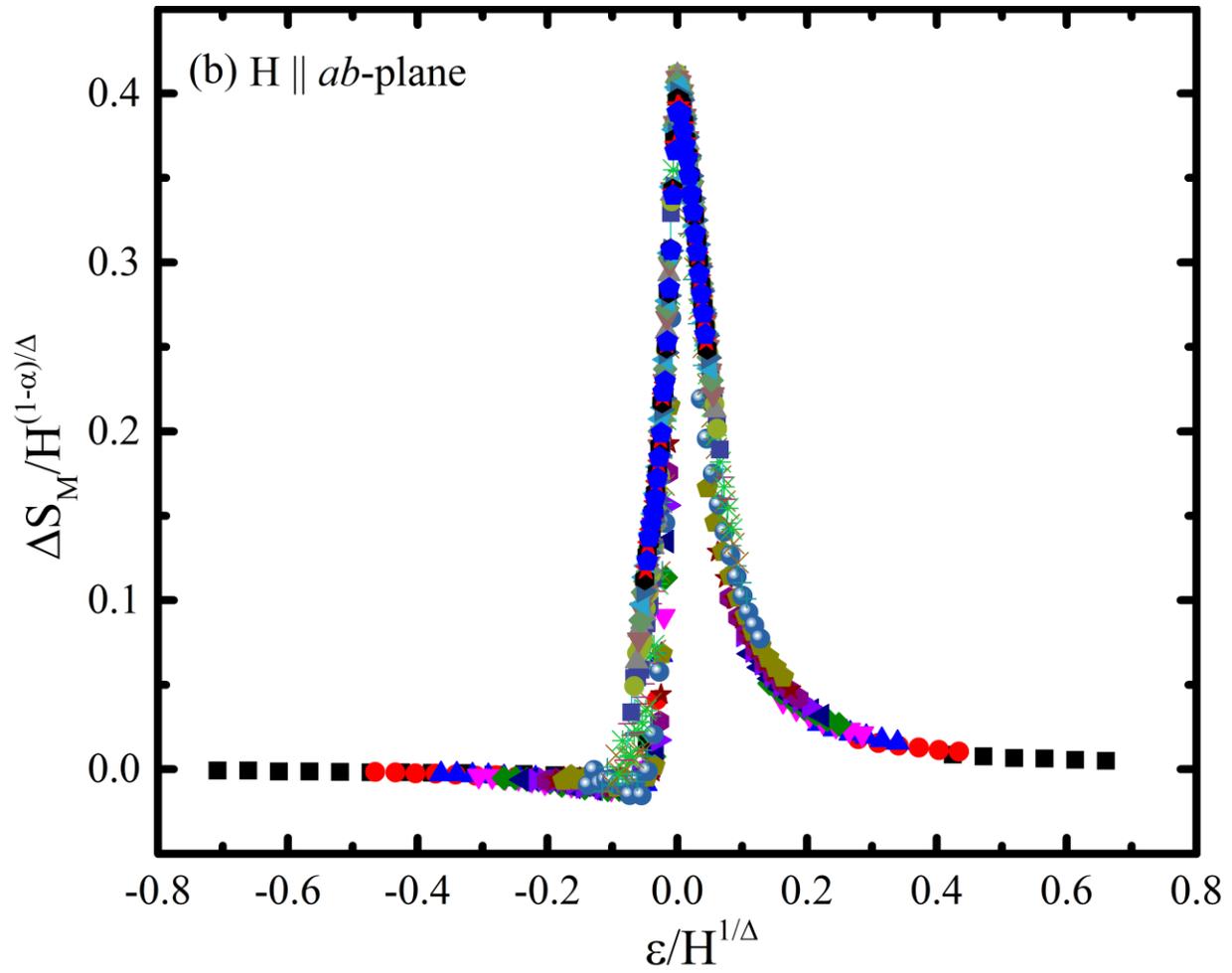

Figure 11: The scaling of the magnetic entropy curves: $-\Delta S_M / H^{(1-\alpha)/\Delta}$ vs $\varepsilon/H^{1/\Delta}$ along the (a) *c*-axis and (b) *ab*-plane.

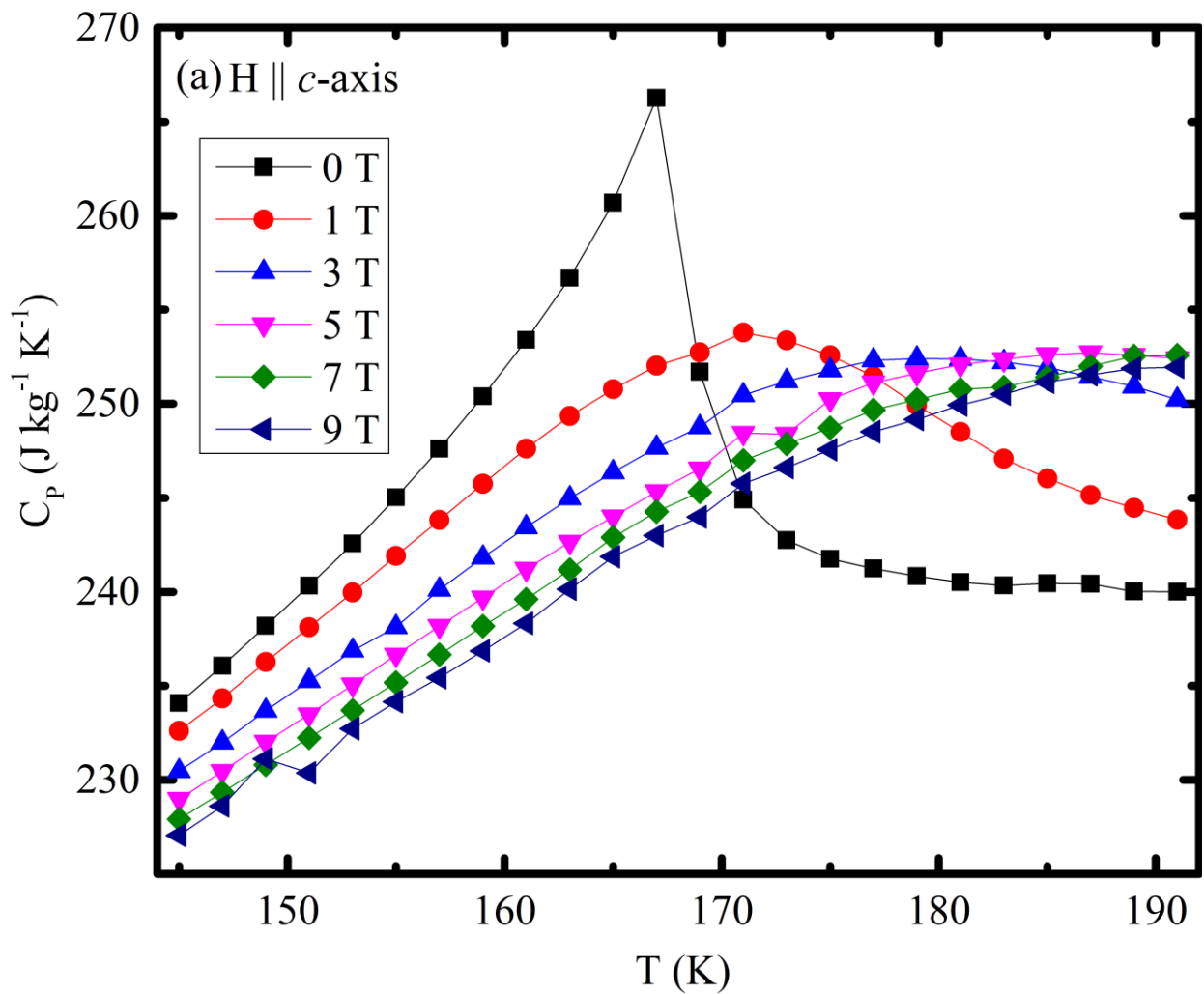

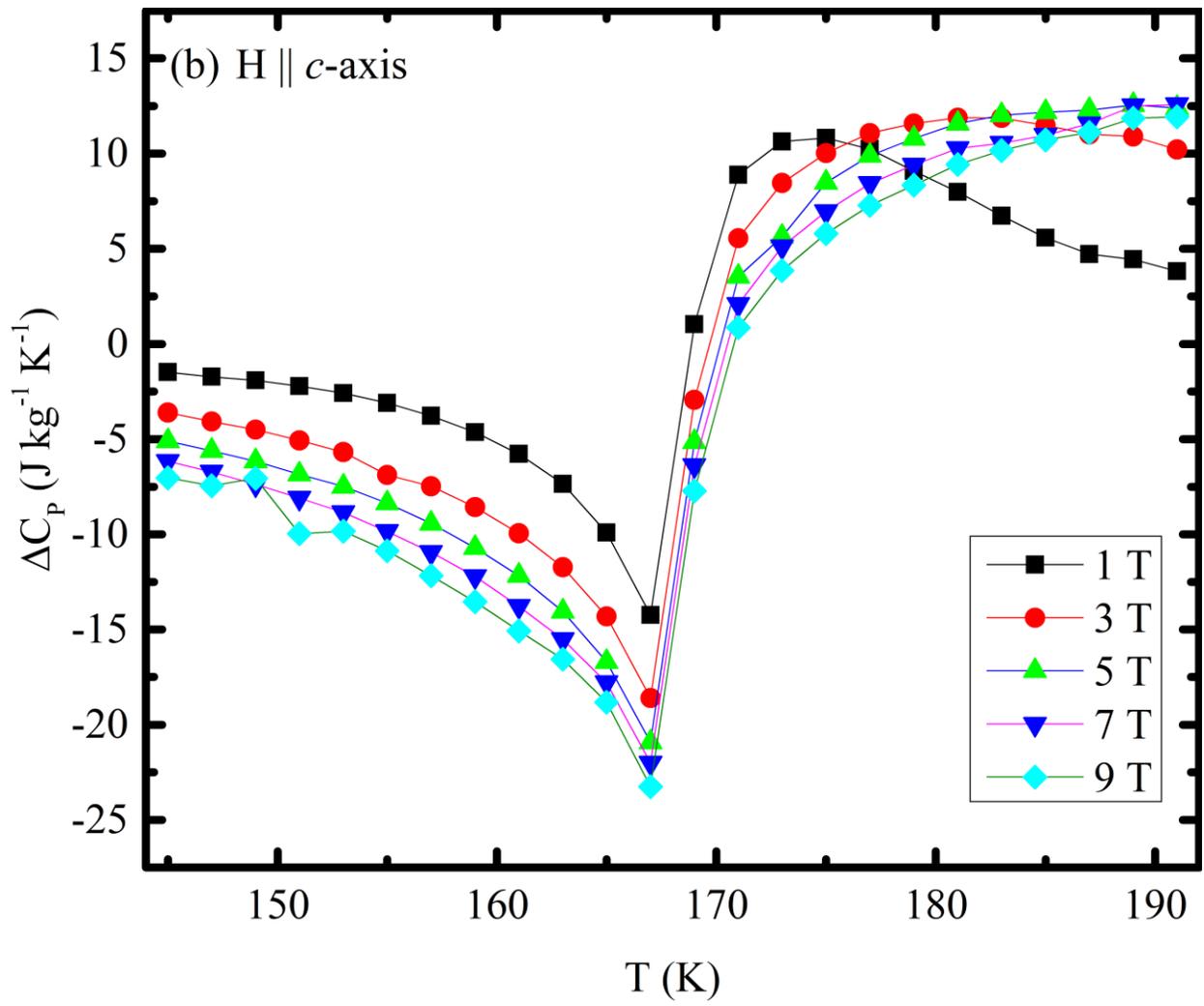

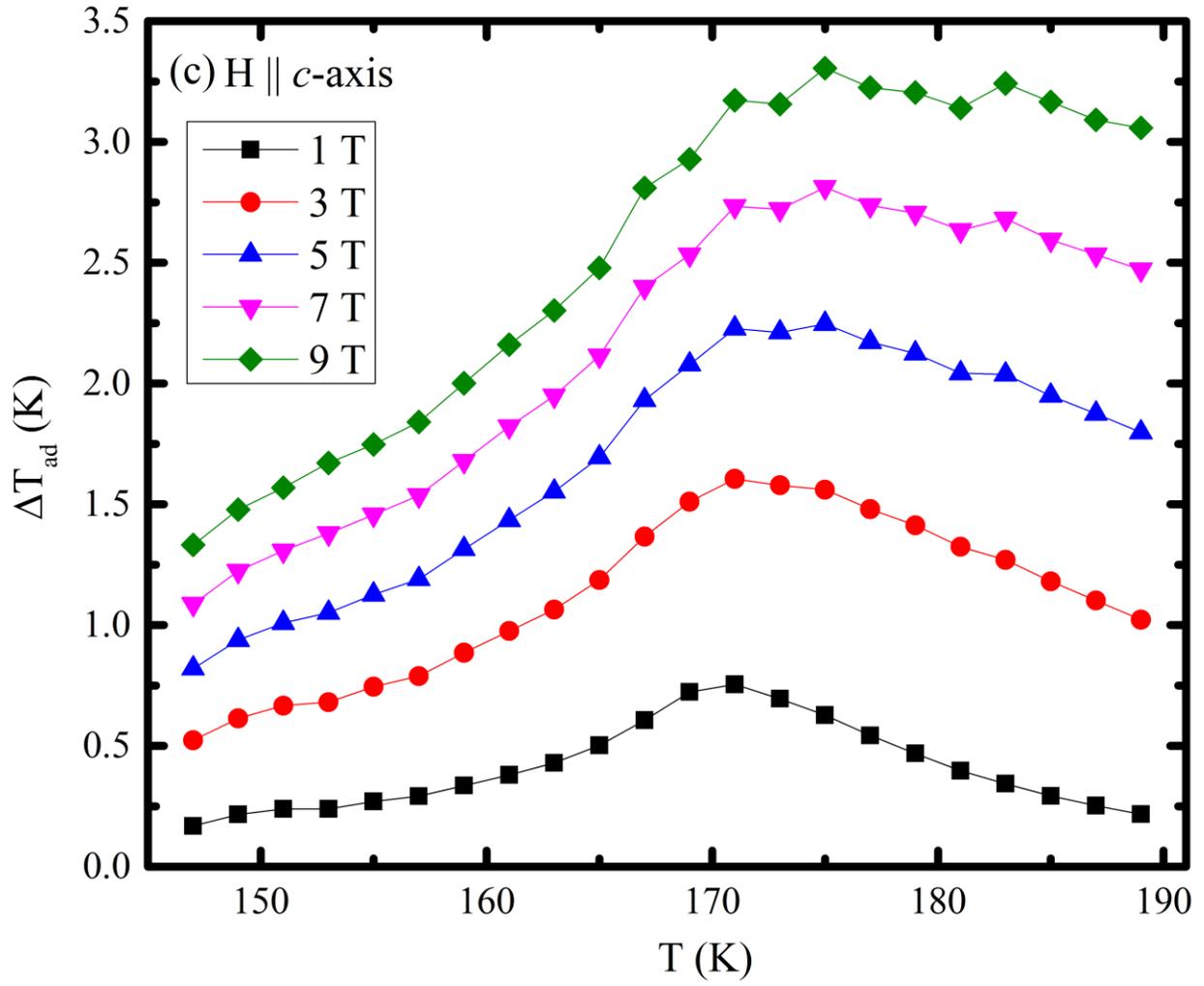

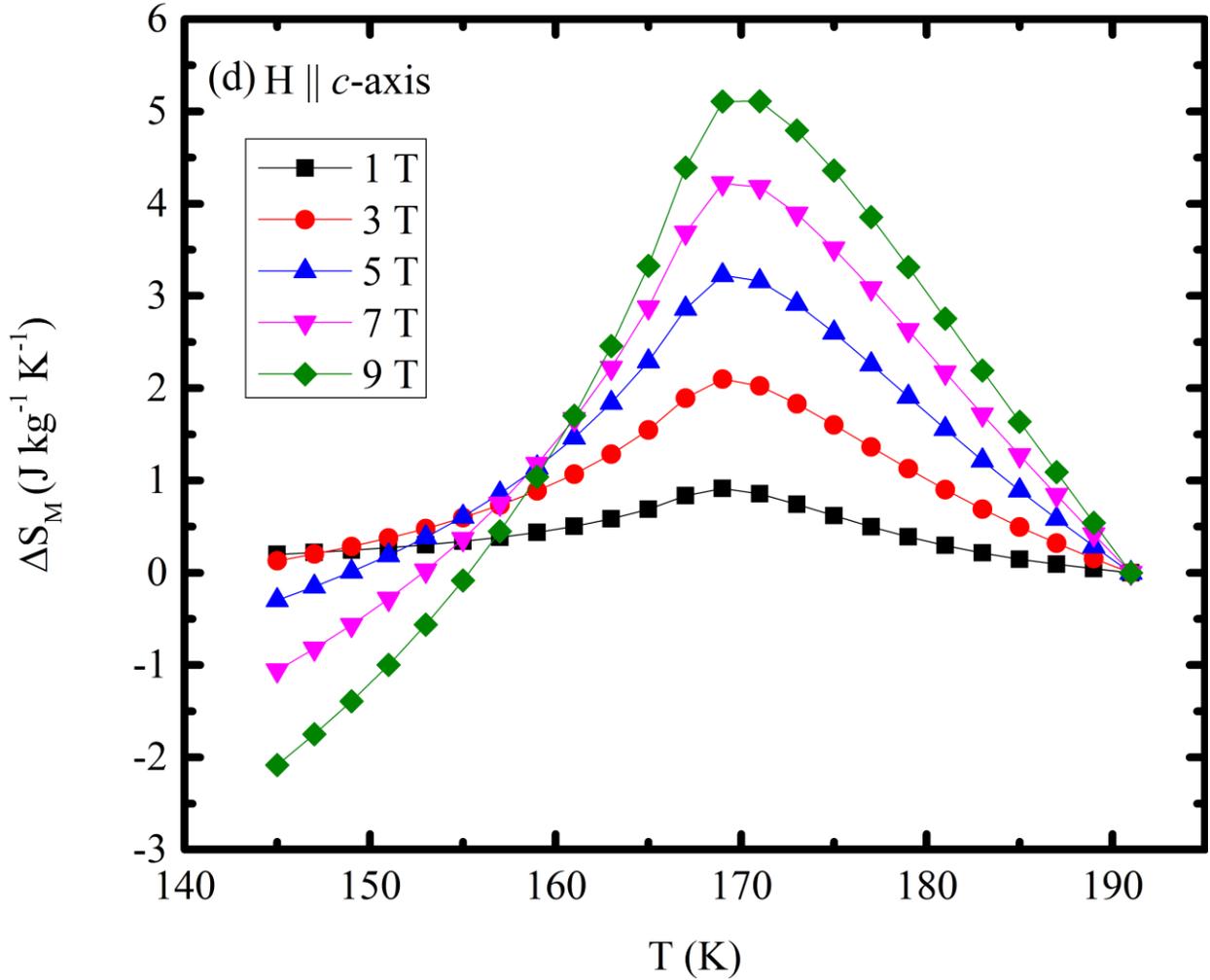

Figure 12: (a) Temperature dependence of specific heat at the indicated magnetic field along the $c$-axis. (b) Temperature dependence of specific heat change [$\Delta C_p = C_p(T, H) - C_p(T, 0)$] at the indicated magnetic field along the $c$-axis. (c) Temperature dependence of adiabatic temperature change $\Delta T_{ad}$ for single crystal $Cr_2Te_3$ estimated from heat capacity data for different magnetic fields. (d) Temperature dependence change of magnetic entropy estimated from heat capacity in different magnetic fields along the $c$-axis.